\documentclass{aa}
\usepackage{natbib}

\AddToHook{begindocument/before}{\RequirePackage{hyperref}}
\usepackage[colorlinks=true, linkcolor=blue, citecolor=blue, filecolor=blue, urlcolor=blue]{hyperref}
\usepackage{graphicx}
\usepackage{epstopdf}
\usepackage{multirow} 

\usepackage{txfonts}
\def\JASTP{\ref@jnl{Journal of Atmospheric and Solar-terrestrial Physics}}

\DeclareUnicodeCharacter{2212}{-}

\begin{document} 

   \title{Evolution of interacting coronal mass ejections driving the great geomagnetic storm on 10 May 2024}
   \author{Soumyaranjan Khuntia\orcid{0009-0006-3209-658X}
          \inst{1,2},
         % \and
          Wageesh Mishra\orcid{0000-0003-2740-2280}\inst{1,2}
          \and
          Anjali Agarwal\orcid{0009-0007-4956-5108}\inst{1,2}
          }
          
   \institute{Indian Institute of Astrophysics, II Block, Koramangala, Bengaluru 560034, India\\
              \email{soumyaranjan.khuntia@iiap.res.in}
              \and
              Pondicherry University, R.V. Nagar, Kalapet 605014, Puducherry, India
             }
   \date{}

\titlerunning {A great geomagnetic storm on 10 May 2024}

\abstract
   {The arrival of a series of coronal mass ejections (CMEs) at the Earth resulted in a great geomagnetic storm on 10 May 2024, the strongest storm in the last two decades.}
  % aims heading (mandatory)
   {We investigate the kinematic and thermal evolution of the successive CMEs to understand their interaction en route to Earth. We attempt to find the dynamics, thermodynamics, and magnetic field signatures of CME-CME interactions. Our focus is to compare the thermal state of CMEs near the Sun and in their post-interaction phase at 1 AU.}
  % methods heading (mandatory)
   {The 3D kinematics of six identified Earth-directed CMEs were determined using the graduated cylindrical shell (GCS) model. The flux rope internal state (FRIS) model is implemented to estimate the CMEs' polytropic index and temperature evolution from their measured kinematics. The thermal states of the interacting CMEs are examined using the in-situ observations from the Wind spacecraft at 1 AU.}
  % results heading (mandatory)
   {Our study determined the interaction heights of selected CMEs and confirmed their interaction that led to the formation of complex ejecta identified at 1 AU. The plasma, magnetic field, and thermal characteristics of magnetic ejecta (ME) within the complex ejecta and other substructures, such as interaction regions (IRs) within two ME and double flux rope-like structures within a single ME, show the possible signatures of CME-CME interaction in in-situ observations. The FRIS-model-derived thermal states for individual CMEs reveal their diverse thermal evolution near the Sun, with most CMEs transitioning to an isothermal state at 6-9 $R_\odot$, except for CME4, which exhibits an adiabatic state due to a slower expansion rate. The complex ejecta at 1 AU shows a predominant heat-release state in electrons, while the ions show a bimodal distribution of thermal states. On comparing the characteristics of CMEs near the Sun and at 1 AU, we suggest that such one-to-one comparison is difficult due to CME-CME interactions significantly influencing their post-interaction characteristics.}
   {}


\maketitle

\section{Introduction}

Coronal mass ejections (CMEs) are dynamic eruptions from the Sun that release enormous quantities of magnetized plasma into interplanetary space \citep{Hundhausen1984,Webb2012,Temmer2023,Mishra2023}. When CMEs travel farther from the Sun within interplanetary space, they are traditionally known as Interplanetary coronal mass ejections (ICMEs). These solar events can cause prolonged geomagnetic storms, disrupting essential societal infrastructure such as communication systems, power grid systems, and satellite operations, and pose significant risks to our technology-dependent society \citep{Gonzalez1994,Pulkkinen2007}. Therefore, two of the primary research objectives in the CME-related field are to forecast the arrival time and evaluate the impact at Earth.

The initial speed of a CME, ranging from 100 to 3000  km s$^{-1}$ \citep{Yashiro2004} within 30 $R_\odot$, can be decided based on the maximum energy that is available from an active region \citep{Gopalswamy2005aa}. However, by the time most CMEs reach Earth, their speeds have reduced to typically around 500 - 600 km s$^{-1}$ \citep{Richardson2010}. Furthermore, the kinematics, thermodynamics, radial expansion magnetic properties, and geoeffective parameters of a CME can also be changed during its evolution as it propagates away from the Sun \citep{Liu2006,Kilpua2017,Mishra2020,Mishra2021a,Khuntia2023,Agarwal2024}. This suggests that while the active region near the Sun influences how powerful a CME can be, its evolution through interplanetary space is crucial in determining its final impact on Earth. Furthermore, the interaction of a CME with another CME or pre-conditioned ambient medium can significantly influence its plasma parameters, arrival time, and geo-effectiveness \citep{Liu2014, Lugaz2017, Desai2020, Mishra2021, Koehn2022, Temmer2023}. There have been extensive case studies on interacting CMEs in the interplanetary space to better understand the changes in their morphology, arrival time, and consequences \citep{Wang2003,Mishra2014,Temmer2014, Mishra2015, Mishra2017, Scolini2020}. However, the exact role of CME-CME interactions in governing the formation of merged ejecta, distinct structures, and their thermodynamic evolution in interplanetary space is still not fully understood.

\setlength{\tabcolsep}{3pt}
\begin{table*}
\caption{\label{tab:GCS} The list of responsible CMEs causing the great geomagnetic storm on 10 May 2024.}
%\hspace{-45pt]
\centering
\begin{tabular} {llcc|ccccccc|c}
\hline
Events & Date, Time & Flare/ & Source  & Time (UT) & Height ($R_\odot$)  & Longitude & Latitude & Aspect  & Tilt  & Half &Max Speed\\

& (UT) & filament & region &Initial-Final & Initial-Final & (deg) & (deg) & Ratio & Angle & Angle & (km s$^{-1}$)\\
 
 & & & & & & & & (deg) & (deg) &\\
\hline
\hline
CME1 & 8, 5:36 & X1.0 & S18W10& 05:48-10:18  & 5.8-26.9 & 16$\pm 3$ & -8$\pm 2$ & 0.27$\pm 0.1$ & 84$\pm 6$ & 24$\pm 3$ & 967\\
\hline
CME2 & 8, 12:24 & M8.7& S19W11  & 12:36-15:54 & 5.4-22.4 & 13$\pm 2$ & -16$\pm 4$ & 0.34$\pm 0.1$  & 27$\pm 5$ & 23$\pm 6$ & 1142\\
\hline
CME3 & 8, 19:12 & Filament & N25E14 & 19:36-23:42 & 5.2-24.7 & -27$\pm 5$ & 4$\pm 6$ & 0.24$\pm 0.1$ & 79$\pm 2$ & 15 $\pm 5$ & 991 \\
\hline
CME4 & 8, 22:36 & X1.02 & S20W14 & 23:06-25:54 & 8.9-28.3 & 6$\pm 7$ & -18$\pm 3$ & 0.26$\pm 0.12$ & 15$\pm 10$ & 18 $\pm 7$& 1406 \\
\hline
CME5 & 8, 22:36 & M9.87 & S19W28  & 23:06-26:42  &  4.6 -22.5  & 38$\pm 6$ & -15$\pm 3$ & 0.15$\pm 0.4$ & -83$\pm 8$ & 16$\pm 8$ & 1103 \\
\hline
CME6 & 9, 9:24 & X2.2 & S20W22 & 09:24-12:06 & 3.9-25.4 & 27$\pm 3$ & -14$\pm 2$ & 0.28$\pm 0.1$ & -77$\pm 4$ & 23$\pm 3$ & 1746\\
\hline

\hline
\end{tabular}

\tablefoot{The second column shows the date and time for the first appearance of each CME in the SOHO/LASC0-C2 field of view. The GCS-model-fitted parameters for the CMEs are shown in the 5$^{th}$ - 12$^{th}$ column. The fourth and fifth columns show the time and height range for which the GCS model fit was done. The last column shows the estimates of the maximum LE speed ($v$) of each CME during our observation duration in the coronagraphic field of view. }

\end{table*}

Understanding the kinematic and thermodynamic parameters of CMEs near the Earth is important as the magnetic reconnection between the southward-directed CME's magnetic field and Earth’s northward-pointing magnetic field facilitates the transfer of energy and plasma within Earth’s magnetosphere \citep{Dungey1961,Tsurutani1988}. The intensity of geomagnetic storms is often represented by the disturbance storm time (Dst) index \citep{Nose2015}, which quantifies the perturbation in the horizontal component of the geomagnetic field at equatorial latitudes. A geomagnetic storm is classified as great if it has reached a minimum Dst index value of -350 nT or lower \citep{Gonzalez2011,Mishra2024}. Since the space age around 1957, there have only been 11 cases when the Dst index exceeded the minimum of −350 nT \citep{Meng2019}. The most intense geomagnetic storm recorded by the Dst index is the March 1989 storm, which reached a minimum Dst index of −589 nT. This caused significant space weather hazards, including a blackout of the Canadian Hydro-Québec power system \citep{Boteler2019}. A great storm occurred in November 2004, with a minimum Dst index value of -373 nT \citep{Yermolaev2008}. Before this, two severe storms took place in October 2003, registering minimum Dst index values of -363 nT and -401 nT, followed by another major storm in November 2003 with a minimum Dst index value of -472 nT \citep{Gopalswamy2005aa,Srivastava2009}. Such great geomagnetic storms are very rare; thus, they pose a significant challenge in statistically analyzing the properties of their drivers. Therefore, it is important to analyze the individual great storms as they happen and examine their drivers if they exhibit significantly different properties.

Recent studies have reported various impacts of this great geomagnetic storm, such as an increase in dayside ionospheric Total Electron Content (TEC), low latitude aurora, radio blackout, and satellite drag, using both ground stations and space satellites \citep{Gonzalez-Esparza2024, Hajra2024, Hayakawa2024, Lazzus2024, Spogli2024, Parker2024, Jain2025}.\citet{Weiler2025} studied the kinematics of the responsible CMEs and demonstrated that sub-L1 missions would have been able to effectively predict the strength of this geomagnetic storm up to 2.57 hours in advance, especially for strongly interacting events, which are still difficult to forecast. \citet{Liu2024} called it a "perfect storm" and highlighted contrasting magnetic fields and geo-effectiveness of the complex structures at two different vantage points, even with only a mesoscale separation between them.

In this letter, we analyze the candidate CMEs driving the great geomagnetic storm, the largest one in the last two decades, which started on 10 May 2024 and reached a minimum Dst index of -406 nT on 11 May 2024. In Section \ref{sec:nearsun}, we examine the near-Sun observations, where we estimate the 3D kinematics of successive CMEs associated with the great storm, focusing on their potential interactions en route to Earth. We also derived the near-Sun thermodynamic evolution of these interacting CMEs by combining remote observations with models. In Section \ref{sec:near-earth}, we analyze the in-situ storm observations near Earth and disentangle the associated structures. Also, we studied the thermal state evolution across the complex structure and ambient solar wind. Finally, in Section \ref{sec:conclusions}, we summarize the results, highlighting the key findings from both remote and in-situ observations.

\begin{figure*}[ht]
   \centering
   \includegraphics[scale=0.21]{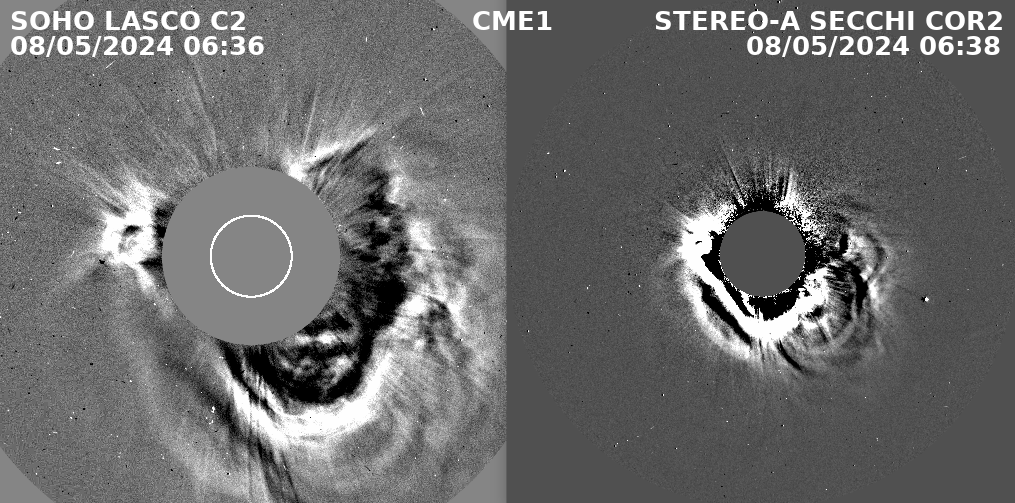}
   \includegraphics[scale=0.21]{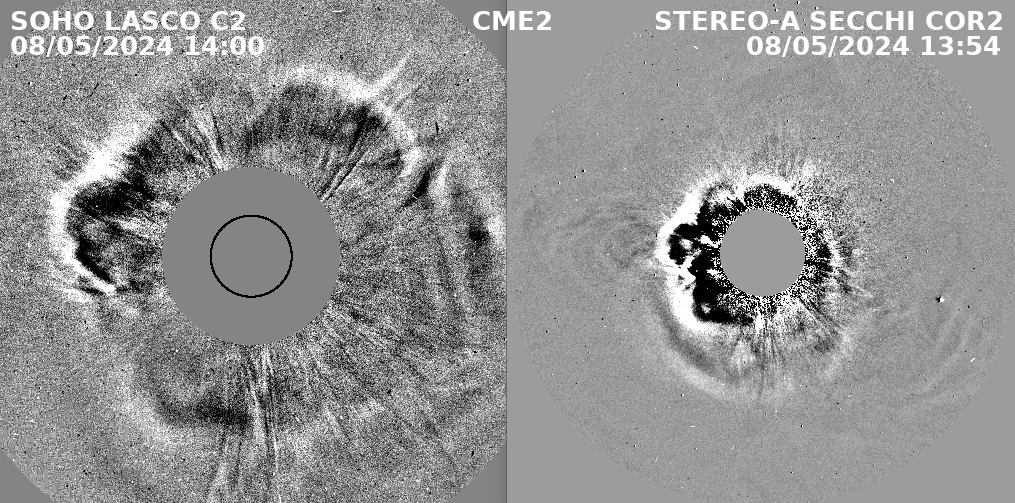}\\ \vspace{-1mm}
   \includegraphics[scale=0.21,trim={0cm 0cm 0cm 2cm}, clip]{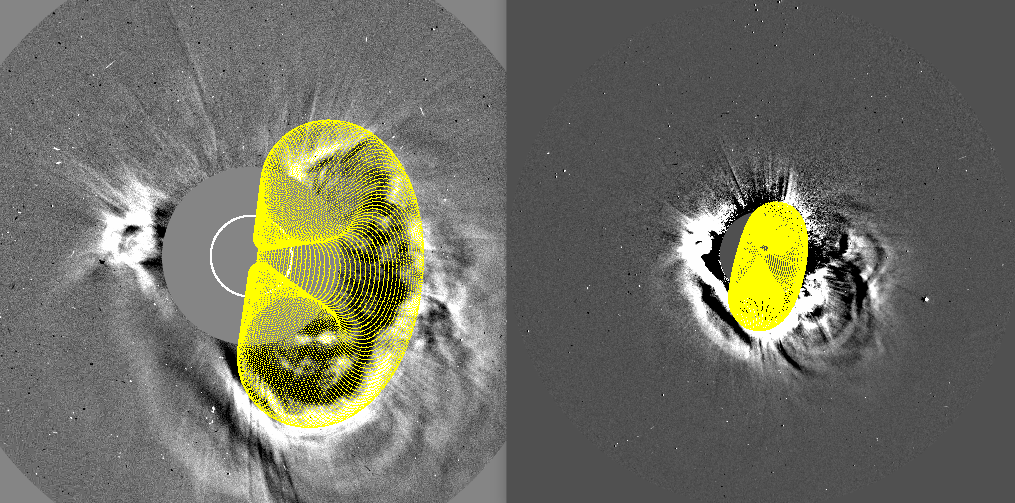}
   \includegraphics[scale=0.21,trim={0cm 0cm 0cm 2cm}, clip]{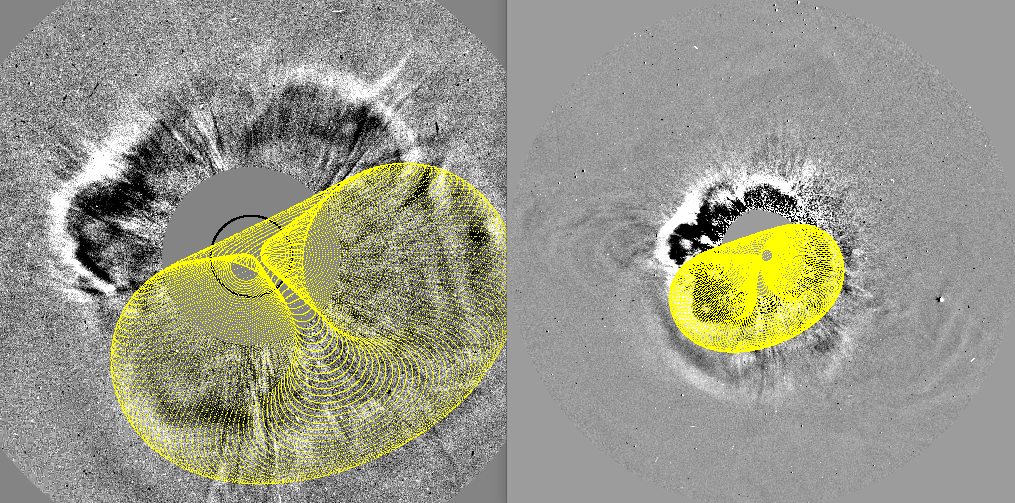}\\
   \includegraphics[scale=0.21]{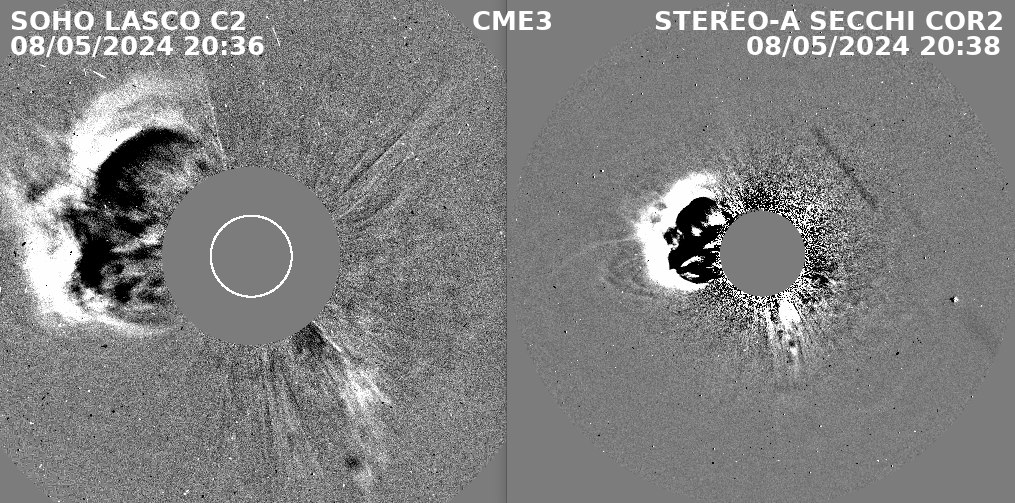}
   \includegraphics[scale=0.21]{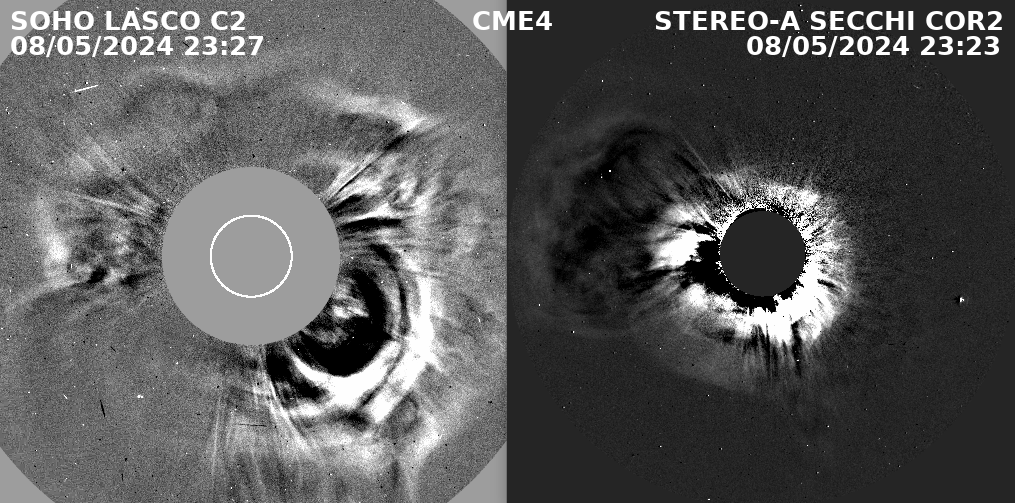}\\
   \vspace{-1mm}
   \includegraphics[scale=0.21,trim={0cm 2cm 0cm 0cm}, clip]{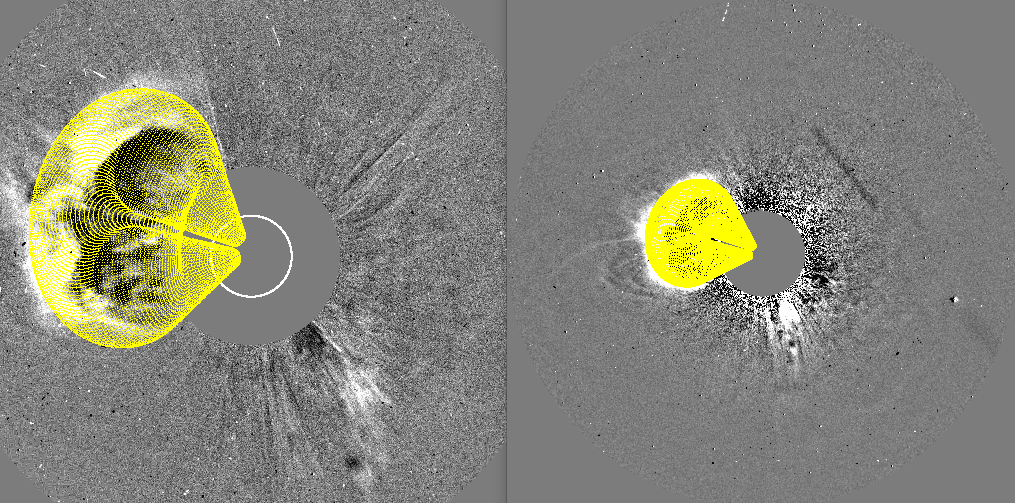}
   \includegraphics[scale=0.21,trim={0cm 0cm 0cm 2cm}, clip]{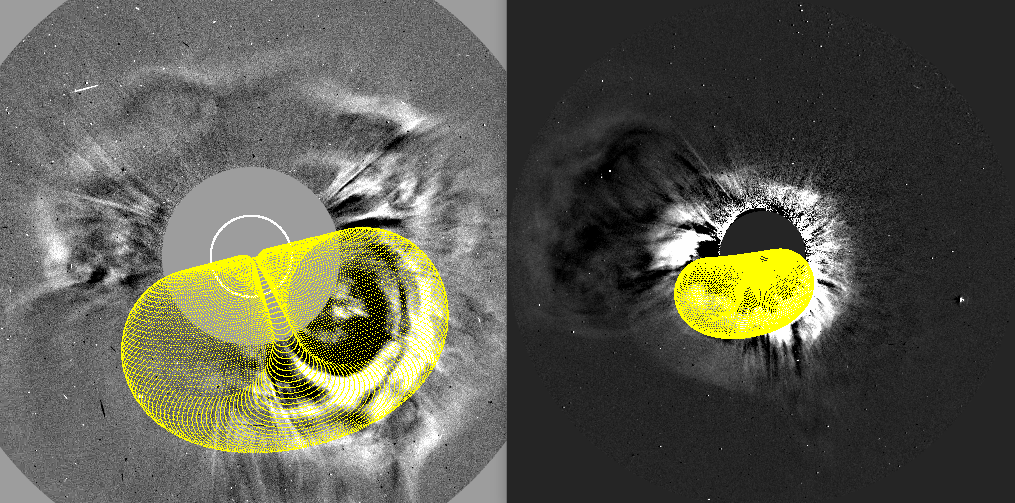}\\
   \includegraphics[scale=0.21,trim={0cm 1cm 0cm 0cm}, clip]{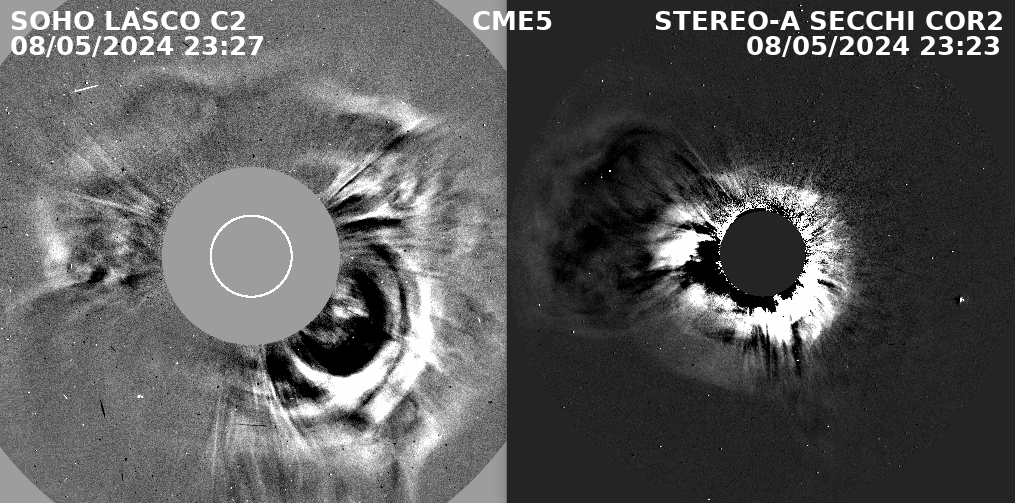}
   \includegraphics[scale=0.21,trim={0cm 1cm 0cm 0cm}, clip]{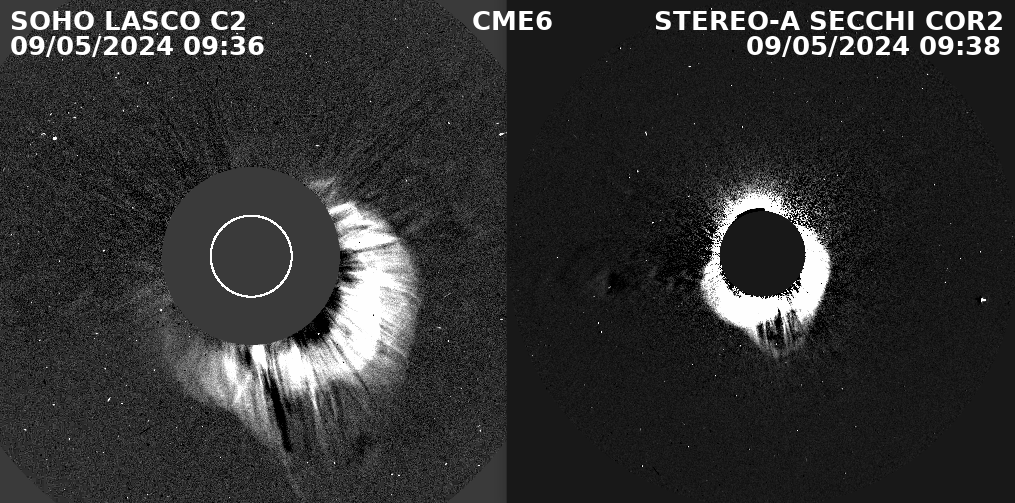}\\ \vspace{-1mm}
   \includegraphics[scale=0.21,trim={0cm 1cm 0cm 2cm}, clip]{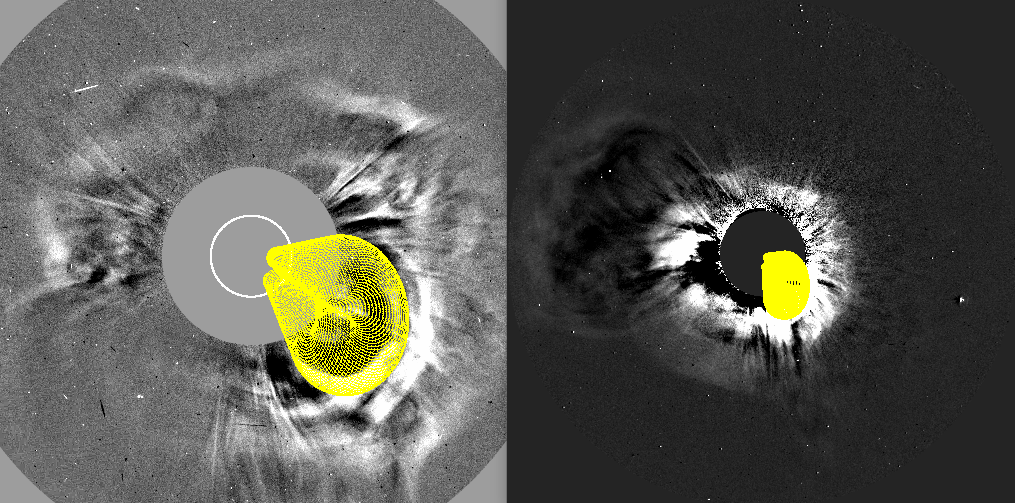}
   \includegraphics[scale=0.21,trim={0cm 1cm 0cm 2cm}, clip]{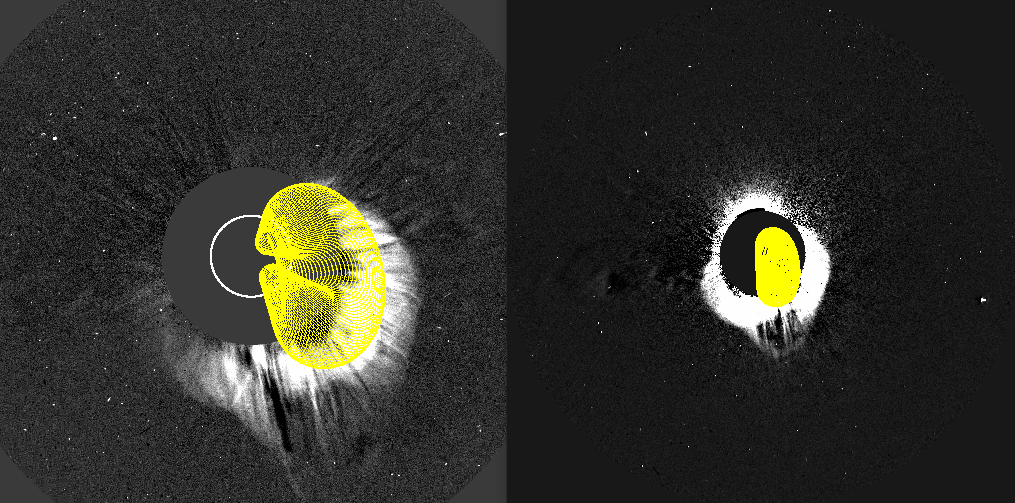}\\
   
   \caption{The GCS model fitted Earth-impacting CMEs, responsible for the great event on 10 May 2024. The top panel of each corresponding CME shows the running-difference imaging observations of CMEs by SOHO/LASCO-C2 and STEREO-A/SECCHI-COR2, whereas in the bottom panel, the fitted GCS model is overlaid as the yellow wire frame.}
   \label{fig:cmes_gcs}
\end{figure*}

\section{Near-Sun observations}{\label{sec:nearsun}}

In May 2024, the Sun experienced significant activity due to the emergence of a complex ($\beta\gamma\delta$) active region formed from the merging of NOAA AR 13664 and 13668. Even after completing a full solar rotation, this active region produced numerous energetic flares. Based on in-situ measured flow speed of $\approx$ 700-1000 km s$^{-1}$ at 1 AU around the onset of the geomagnetic storm on 10 May 2024, we investigated potential halo and partial halo CME ejections from the Sun on 8-9 May 2024. We use coronagraph data from the COR2 telescope on the Sun‐Earth Connection Coronal and Heliospheric Investigation (SECCHI; \citealt{Howard2008}) on board the Solar Terrestrial Relations Observatory Ahead (STEREO‐A; \citealt{Kaiser2008}) and the Large Angle Spectroscopic Coronagraph (LASCO; \citealt{Brueckner1995}) on board the Solar and Heliospheric Observatory (SOHO; \citealt{Domingo1995}), to identify probable Earth‐directed CMEs that might have erupted from the Sun 2-3 days before the start of geomagnetic storm. During our investigation period, STEREO‐A was located $\approx 12^{\circ}$ west of the Sun-Earth line, and it provided a similar view of the CMEs as from the SOHO viewpoint. Furthermore, we search for corresponding activity on the solar disc using extreme ultra‐violet imagery from the Atmospheric Imaging Assembly (AIA; \citealt{Lemen2012}) instrument on board the Solar Dynamics Observatory (SDO; \citealt{Pesnell2012}) to locate the associated source region. By analyzing the source locations and speed profiles, we identified six potential CMEs that could have contributed to the great event, as shown in Table \ref{tab:GCS}. The selected CMEs were ejected on 8-9 May 2024 (The first appearance of CMEs on LASCO/C2 is shown in the second column of Table \ref{tab:GCS}). Furthermore, we associated these CMEs with their corresponding flares and filaments by assessing the timing of the brightest flare or filament and evaluating the likelihood of them being the source. Out of the six CMEs, CME3 is associated with a filament eruption in active region 13667, located at N25E14, while the remaining CMEs are flare-related and originate from active region 13664 (third and fourth column in Table \ref{tab:GCS}).

\subsection{3D kinematics using white-light observations }{\label{subsec:kinematics}}

We have applied the Graduated Cylindrical Shell (GCS) model \citep{Thernisien2006, Thernisien2011} to determine the 3D leading edge (LE) height (h) and direction of the CMEs. The GCS model provides a simplified geometric framework representing the CME's magnetic flux rope topology in space. This model has been used routinely to mitigate the observed projection effect in CME kinematics \citep{Liu2010, Wang2014, Mishra2015}. This model takes advantage of fitting the CME envelope with simultaneous observations from multiple vantage points to obtain a better understanding of the CME kinematics. Our work used the contemporaneous coronagraphic observations from STEREO-A/COR2 and SOHO/LASCO-C2 \& C3. The GCS-fitted coronagraphic images are shown in Figure \ref{fig:cmes_gcs}.

The LE heights and geometrical \& positional parameters obtained from the GCS model fitting of the selected CMEs are mentioned in Table \ref{tab:GCS}. CME3, with the GCS model-derived source location of N04E24, erupted from the northeast side of the solar disk, while all other CMEs erupted from the southwest side. We found no substantial change in tilt angle, aspect ratio, and half angle within our observed field of view. The errors mentioned in Table \ref{tab:GCS} for the GCS-fitted parameters were calculated by repeating the fitting process multiple times. Moreover, it is challenging to quantify the error associated with the GCS model, as the fitting process is performed manually and depends on the user's experience and interpretation of the event. \citet{Thernisien2009} estimated the mean errors in the GCS model to be approximately $\pm 4.3 ^\circ$, $\pm 1.8 ^\circ$, $\pm 22 ^\circ$,  $_{-7^\circ}^{+13^\circ}$, $_{-0.04^\circ}^{+0.07^\circ}$, $\pm 0.48 R_\odot$ for longitude, latitude, tilt angle, half angle ($\alpha$), aspect ratio ($k$) and LE height, respectively.

\begin{figure*}[ht]
   \centering
   \includegraphics[scale=0.22,trim={5cm 1cm 10cm 0cm}, clip]{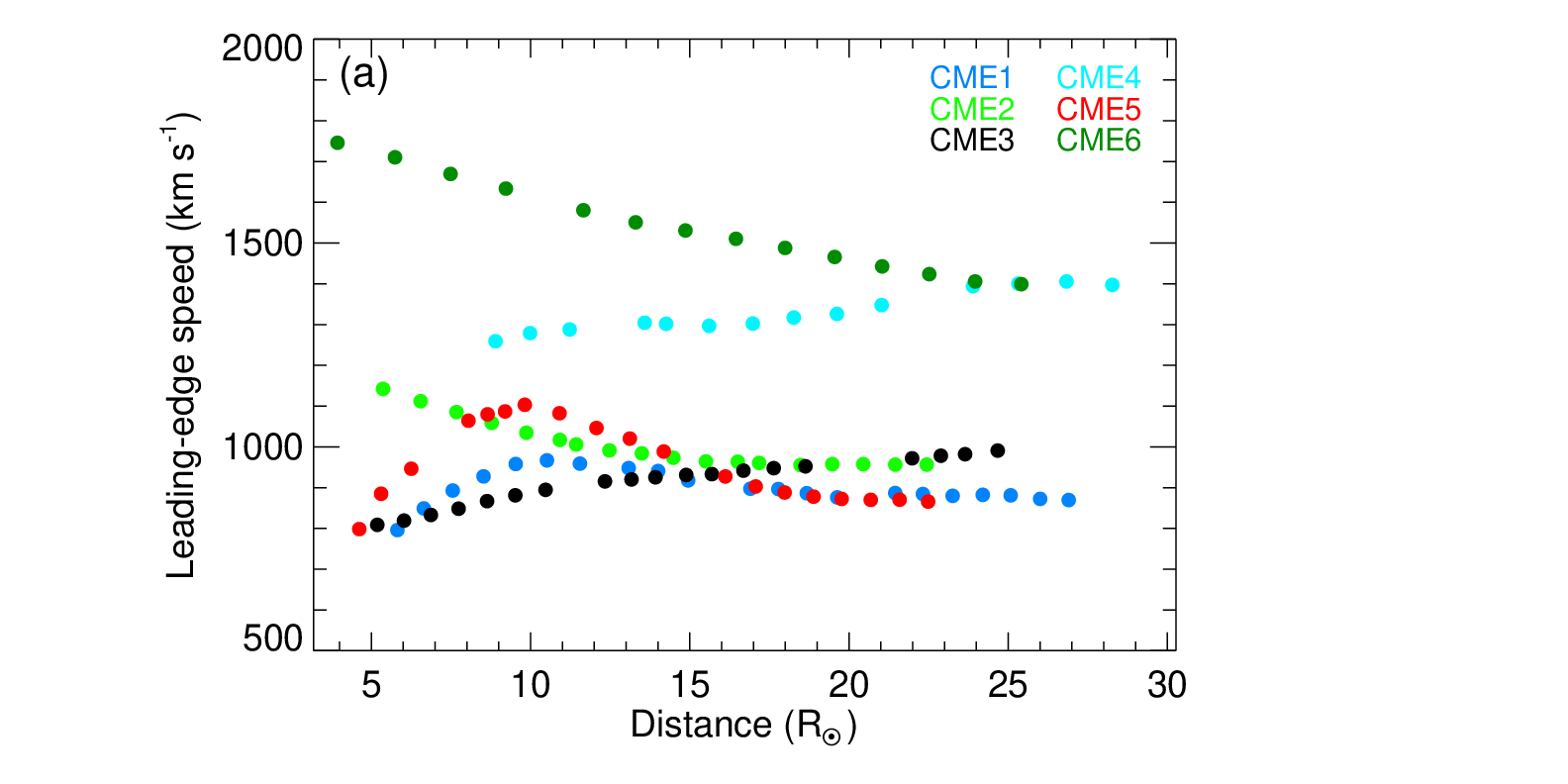}
   \includegraphics[scale=0.22,trim={5cm 1cm 10cm 0cm}, clip]{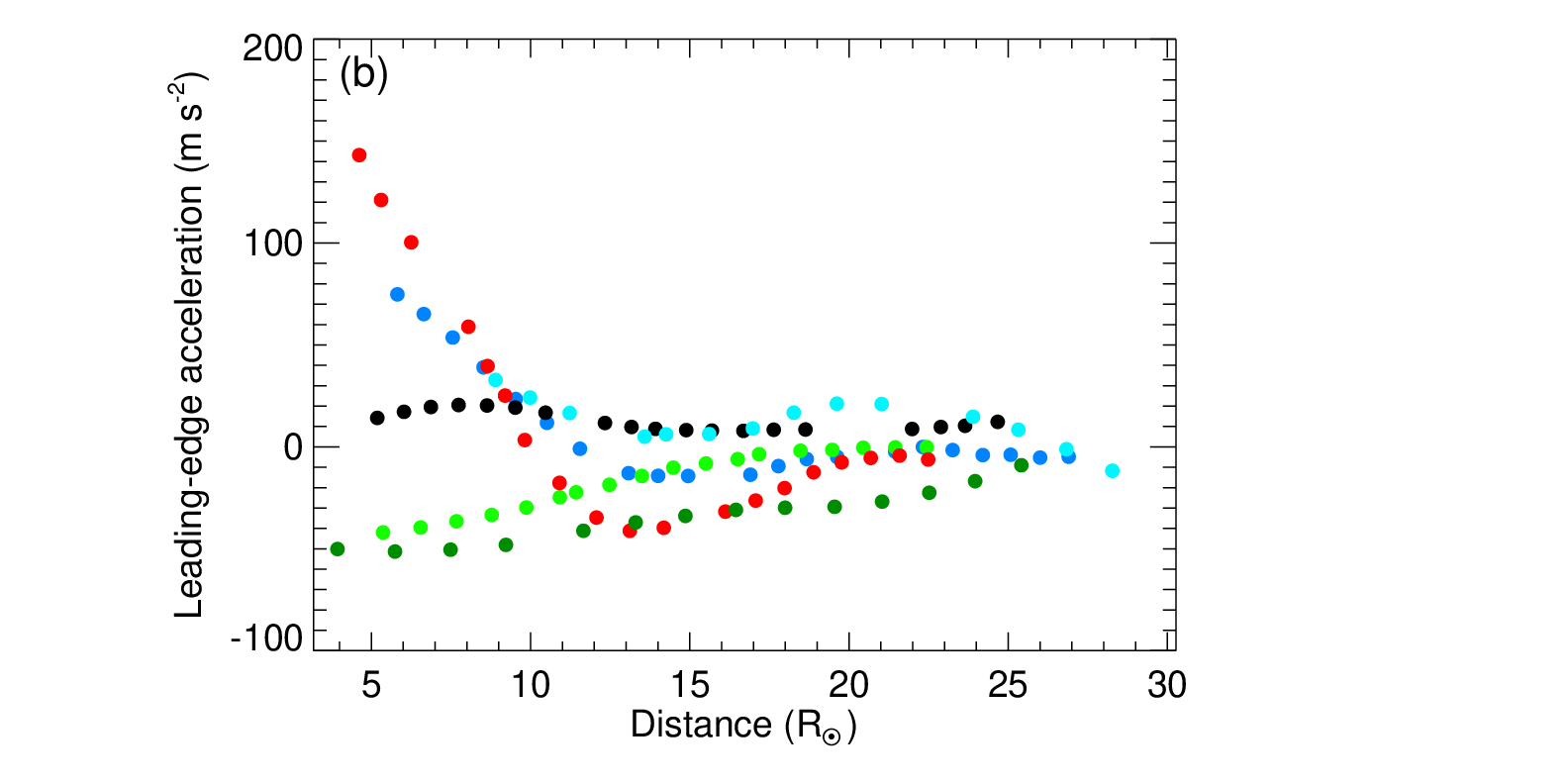}\\
    \includegraphics[scale=0.58]{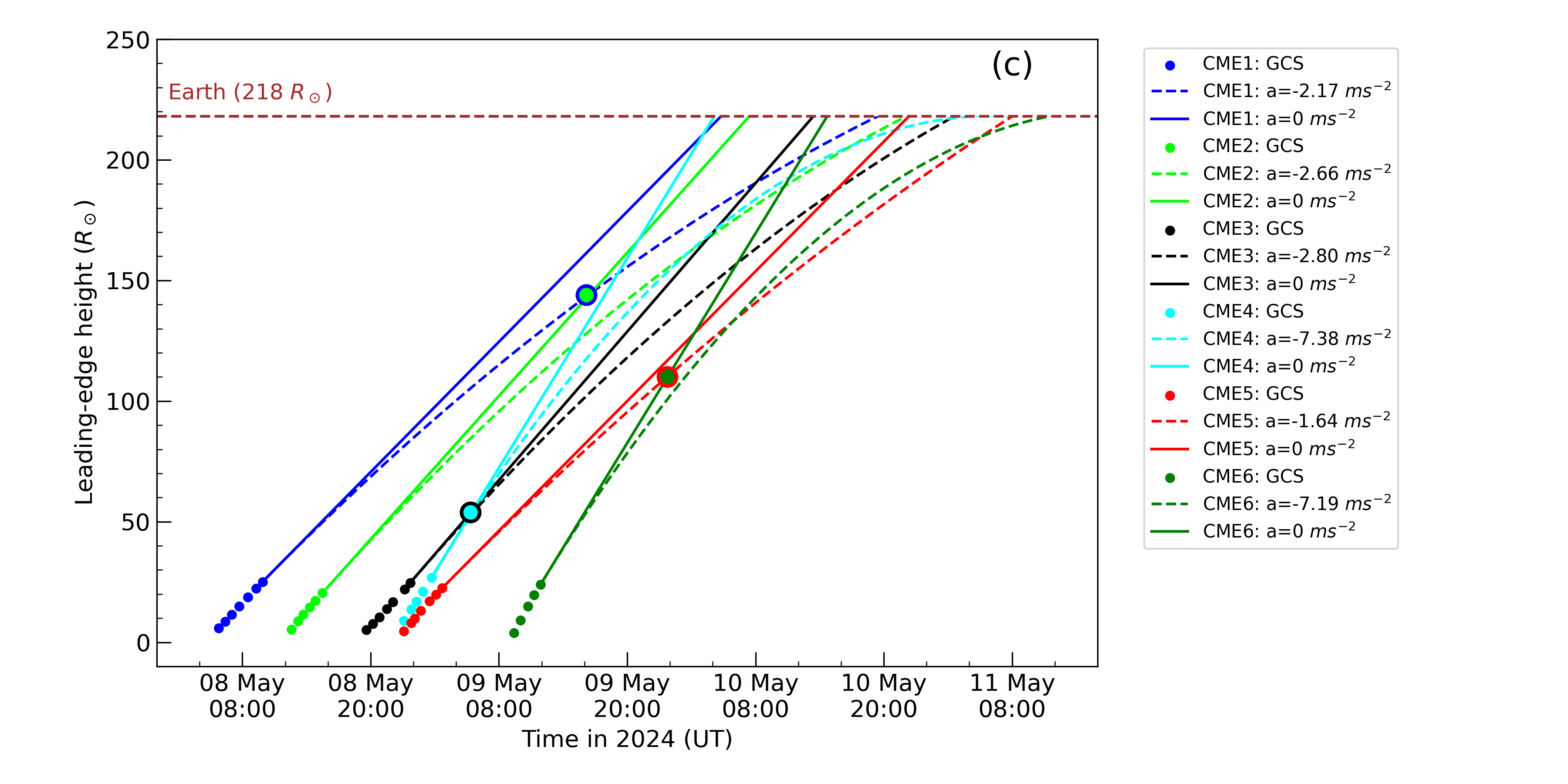}
      \caption{Propagation (a) speed and (b) acceleration of LE of the selected CMEs derived from the GCS model fitted 3D LE heights. (c) The height-time profiles of the CMEs by considering no acceleration (solid line) and some value of constant average deceleration (dash line). The colored-filled circles indicate the first possible CME-CME interaction height. The initial dots show the GCS-fitted heights for the CMEs. We skipped two alternate data points while plotting the GCS-fitted heights for the best visual purpose only. }
         \label{fig:kine}
\end{figure*}

The LE speed ($v$) and acceleration ($a$) of the selected CMEs (Fig. \ref{fig:kine}) are calculated by doing successive time derivatives of the measured 3D LE height ($h$). We applied a moving three-point window, and a linear fit was used within each window to compute the time derivatives at the middle point of the window. For the endpoints (first and last points), derivatives were determined similarly using a two-point window. This method, as described in detail in \citet{Agarwal2024}, allows us to accurately capture reasonable fluctuations in speed and acceleration without decreasing the number of data points in the derivatives. The last column in Table \ref{tab:GCS} shows the maximum speed derived for each CME using the GCS model in the observed field of view. Furthermore, to quantify the error in the LE height, we considered the CME’s sharp LE near the Sun and its more diffuse edge at greater heights. Through multiple fitting attempts, we estimated a maximum uncertainty of $\pm 10\%$ for the LE height, and this error was propagated to derive the kinematic parameters, such as speed and acceleration. Figure \ref{fig:kine}a shows interesting findings that the last CME, CME6, has a higher speed compared to the others, suggesting a potential interaction at greater heights.

Furthermore, we have extrapolated the GCS model derived 3D LE height for each CME using a constant acceleration up to a height of 218 $R_\odot$. The extrapolation of CME height is carried out using the following two approaches: (i) We choose the acceleration to be 0, i.e., the CMEs are propagating with constant speed beyond the last tracked height of LE from the GCS model. (ii) we use the equation of motion with constant acceleration, i.e., $s=ut+{at^2}/2$, where $u$ is the speed of CME LE at the last tracked height from the GCS model, $s$ is the height difference between the last tracked height and 218 $R_\odot$, and $t$ is the time interval between the arrival of CME LE at the last GCS-model-tracked height and at near-earth in-situ spacecraft.  The arrival of individual CMEs LE at in situ spacecraft is identified based on several plasma and magnetic field parameters, which will be discussed in Section~\ref{sec:near-earth}. The extrapolation, in the case of non-zero acceleration, is to match the estimates of CME arrival time with the observed CME/ME LE arrival time at 1 AU. The approach of extrapolation with zero acceleration would only be valid before the CMEs interacted and cannot provide correct estimates of their arrival at 1 AU. However, extrapolation with zero acceleration, together with derived non-zero acceleration, can help determine the possible height range for CME-CME interactions. In general, fast CMEs reach their peak speed within a height of 10 $R_\odot$ and show most of their acceleration up to 20 $R_\odot$ \citep{Zhang2004, Vrsnak2007, Temmer2010}. This result can also be seen in Figure \ref{fig:kine}b; the LE acceleration for all the CMEs decreases and tends towards a lower value. The CMEs primarily experienced aerodynamic drag at larger heights due to interaction with the solar wind. This drag force tends to accelerate the slow CMEs and decelerate the fast CMEs \citep{Gopalswamy2000, Vrsnak2007}. Thus, we can expect that zero acceleration will serve as an upper bound for kinematics.

Without involving any complex approach for estimating the dynamics of the selected CMEs in a varying ambient solar wind, we derived the kinematics for the CMEs beyond coronagraphic heights by extrapolating the measured kinematics from the GCS model to get the range of possible heights for CME-CME interaction. Obviously, the kinematics of the CMEs after their possible interaction cannot be well represented by the extrapolated kinematic profiles of each CME. It is also possible that different CMEs experience different pre-conditioned mediums and follow varying acceleration profiles during their journey in both the pre-and post-interaction phases \citep{Shen2012,Mishra2017}. However, a simple extrapolation of kinematics can serve the scenario of possible interactions and can help interpret the in-situ observations of structures at 1 AU driving the great geomagnetic storm. Figure \ref{fig:kine}c shows the estimated kinematics for each CME up to 218 $R_\odot$. We note that there is not much difference in the GCS model derived LE speed for both CME1 and CME2 at 22 $R_\odot$ (Fig. \ref{fig:kine}a). However, as CME1 is propagating ahead of CME2, CME1 is likely to get a higher drag and deceleration at higher heights than CME2. Moreover, the source longitude of CME1 and CME2 are 16$\pm$3 and 13$\pm$2, respectively. Hence, there is a possibility that the LE of CME2 will catch up with the back of CME1. Considering the derived acceleration of a=-1.78 m s$^{-2}$ and a=-0.71 m s$^{-2}$ for CME1 and CME2, respectively, they are predicted to interact at height $\approx$ 144 $R_\odot$. There is an early interaction between CME3 and CME4 at a LE height of $\approx$ 54 $R_\odot$ (considering $a=0$ m s$^{-2}$). Considering the longitude of the source region for CME3 and CME4, they are more likely to have a flank interaction. This interaction can also be seen in the LASCO C3 coronagraphic view. Furthermore, the derived kinematics also show the interaction between CME5 and CME6 at $\approx$ 110 $R_\odot$. The interaction is possible primarily considering the faster speed of CME6 than CME5 and their close longitude source regions. If CME6 continues to maintain its high speed even after the interaction (as indicated by the faster speed of the trailing magnetic ejecta observed at 1 AU), it suggests that all contributing CMEs likely interacted with one another en route to Earth. This interaction would have resulted in a large-scale complex magnetic structure to be observed at 1 AU, which will be further discussed in the upcoming sections.

We note that our approach assumes a constant acceleration throughout the CME evolution beyond the coronagraphic heights to get a broad range of distances for possible CME-CME interaction. The estimated heights for different interacting CME pairs can change if we consider the drag force between an individual CME and ambient medium, momentum exchange during interacting CME pairs, and possibly magnetic interaction between closely separated CMEs. A detailed study using Heliospheric Imager (HI) observations and focusing more on the momentum exchange and nature of interaction/collision for interacting CMEs \citep{Mishra2015, Mishra2017}, geoeffectiveness \citep{Lugaz2014, Scolini2020} can reveal more insight into the complex ejecta evolution. The changes in the kinematics of CMEs will have imprints on their thermal state, which we will discuss in the next section.

\begin{figure*}[ht]
   \centering
   \includegraphics[scale=0.22,trim={5cm 1cm 10cm 0cm}, clip]{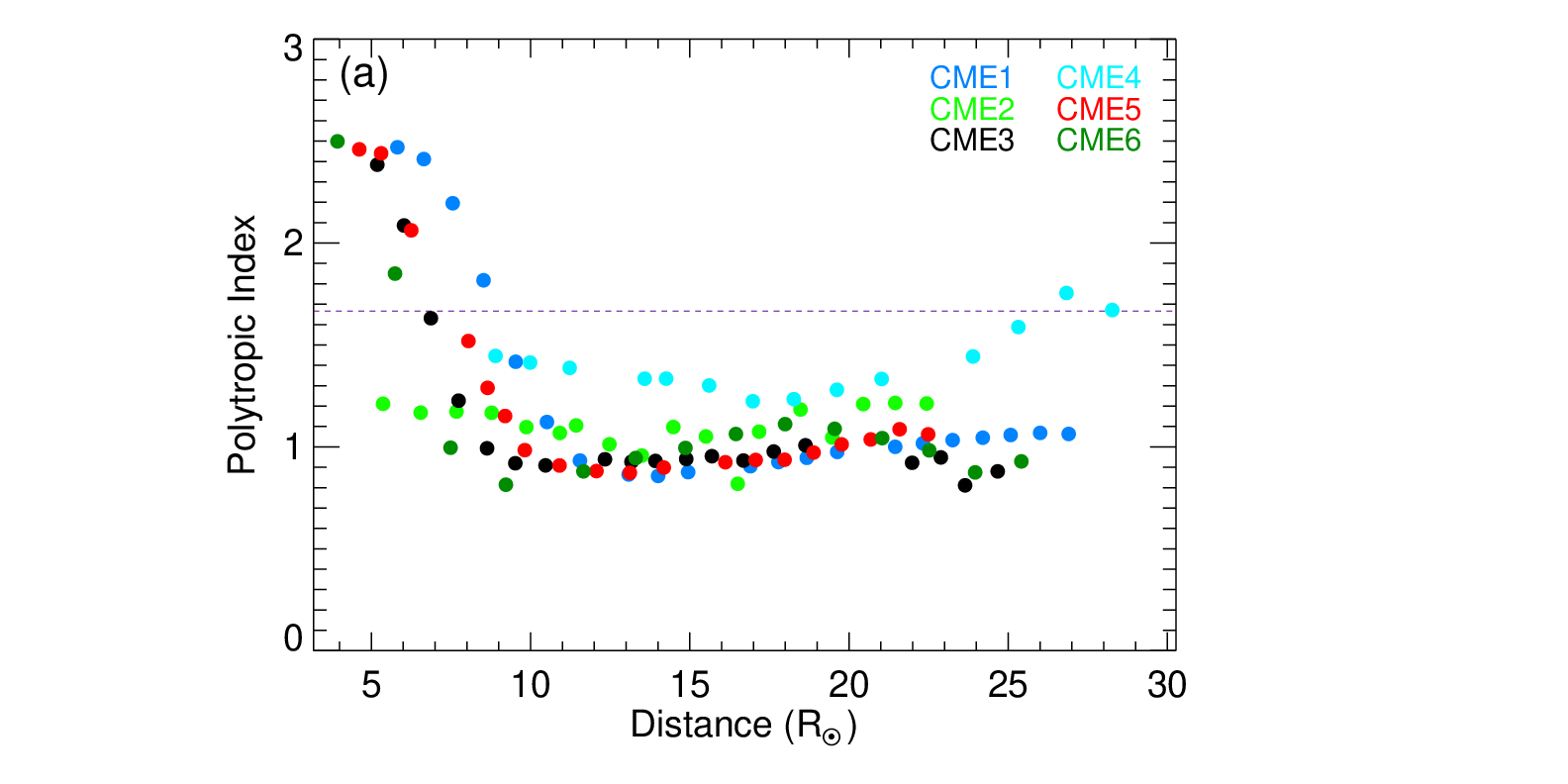}
   \includegraphics[scale=0.22,trim={5cm 1cm 10cm 0cm}, clip]{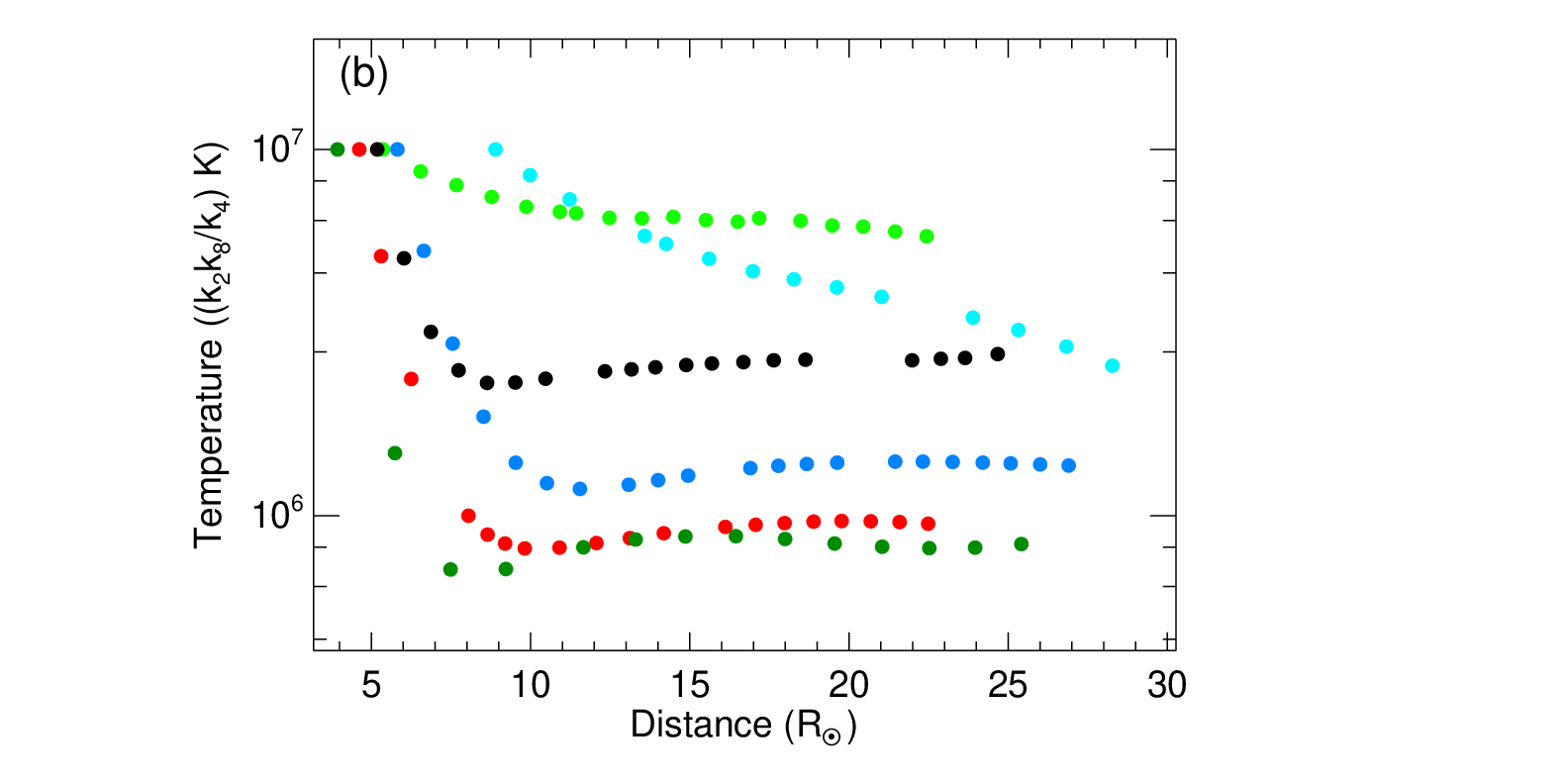}
      \caption{FRIS model-derived (a) polytropic index ($\Gamma$) with the height of the CME LE, the black dashed line represents the value of the adiabatic index ($\gamma$ = 1.67) and (b) temperature (T) profile with the height of the CME LE for the six selected CMEs.}
         \label{fig:gamma}
\end{figure*}

\subsection{Thermodynamics using the measured 3D kinematics}{\label{subsec:thermal}}

Thermodynamics of successively launched CMEs from the same active region is not well understood, and further, there are limited studies to estimate the thermodynamic properties of different substructures of complex ejecta formed due to possible CME-CME interaction before they reach 1 AU. We applied an analytical model, the Flux Rope Internal State (FRIS) Model \citep{Mishra2018, Khuntia2023, Khuntia2024}, to derive the internal thermal properties of the selected CMEs. The model considers the CME to have an axisymmetric cylindrical shape at a local scale and evolve as a polytropic process. The model conserves mass and angular momentum while assuming a self-similar evolution. The model solves the ideal magnetohydrodynamic (MHD) equations of motion for the flux rope, incorporating various internal forces responsible for expanding the CME, such as Lorentz force, thermal pressure force, and centrifugal force. As the model assumes the flux rope plasma is a single species magnetic fluid, the model-derived parameters show the average properties of the CME as a whole, both for protons and electrons. The model uses the global kinematics, such as height and radius of the CME flux-rope, to derive various internal parameters, summarized in Table 1 of \cite{Khuntia2023}.

We have implemented the measured 3D kinematics in the FRIS model using the procedure mentioned in \citet{Khuntia2023}. The equation of motion for the radial expansion of the CME flux-rope can be expressed as,
\begin{align}
\frac {R}{L} = &  { c_5 \biggl[\frac {a_e R^2}{L}\biggr]} - {c_3 c_5 \biggl[\frac {R}{L^2}\biggr]} - {c_2 c_5\biggl[\frac {1}{R}\biggr]} - {c_1 c_5\biggl[\frac {1}{LR}\biggr]}{\nonumber}\\
&+ c_4 \biggl[\frac {da_e}{dt} + {\frac {(\gamma -1)a_e v_c}{L}} + \frac{(2\gamma -1)a_e v_e}{R}\biggr] {\nonumber}\\
&+ c_3 c_4\biggl[ \frac{(2-\gamma) v_c }{L^2 R} + {\frac{(2-2\gamma) v_e }{LR^2}}\biggr] {\nonumber}\\ 
&+ c_2 c_4\biggl[{\frac{(4-2\gamma) v_e L}{R^4}} - {\frac{\gamma v_c}{R^3}}\biggr] {\nonumber}\\
&+ c_1 c_4\biggl[{\frac{(4-2\gamma) v_e}{R^4}} + {\frac{(1-\gamma)v_c}{LR^3}}\biggr]
\label{eqn:fitting1}
\end{align}

\noindent where the inputs to the FRIS model are the distance of the center of the flux-rope from the surface of the Sun (L), the radius of the flux-rope (R), and their successive time derivatives, such as propagation speed ($v_c$) and acceleration ($a_c$) of the axis of the flux rope, expansion speed ($v_e$) and acceleration ($a_e$) of the flux rope. $\gamma$ is the adiabatic index ($\gamma=5/3$ for monoatomic ideal gases), and $c_1-c_5$ are unknown constants coefficients, whose values can be obtained by fitting Equation \ref{eqn:fitting1}. The fitting results for all the six selected CMEs are shown in Appendix (Fig.~\ref{fig:FRIS_error}). Among the several internal properties that can be derived using the FRIS model, for this study, we focus on the evolution of two critical properties, such as the polytropic index ($\Gamma$) and temperature ($T$) of the selected CMEs. The model-derived expression for $\Gamma$ and $T$ are,

\begin{equation}\label{eqn:gamma_eqn}
    \Gamma =  \gamma +   \frac{ln{\frac{\lambda (t)}  {\lambda (t+\Delta t)}}}  {ln[(\frac{L(t+\Delta t)}  {L(t)})[\frac{R(t+\Delta t)}  {R(t)}]^2]} 
\end{equation}

\begin{equation}\label{eqn:temp_eqn}
     {T}=  \frac{ k_2 k_8 }{k_4 } \biggl[ \frac{\pi \sigma}   {(\gamma -1)} {\lambda (LR^2)^{1{-\gamma}}  }\biggr].
\end{equation}

Excluding L, R, and $\gamma$, all other quantities in the above equations are unknown (for details, see \citealt{Khuntia2023}). By determining the fitting coefficients for Equation \ref{eqn:fitting1}, some of those quantities can be evaluated. Thus, apart from $\gamma$, the temperature ($T$) estimates from the model are multiplied by a factor ($ \frac{ k_2 k_8 }{k_4 } $), the absolute value of which could not be derived from the model.
This factor differs for each CME as it depends on the fitted coefficients of individual CMEs but does not change with time for a particular CME. This prevents us from investigating the absolute value for $T$; therefore, we have normalized their relative values to an initial value of $10^7 K$ to compare the temperature changes of different CMEs. The scaling factor is chosen carefully so that the relative temperature values for all the CMEs become equal at the first observed data point. This can enable us to examine the relative change in the trend of temperatures for all the CMEs with distances away from the Sun.

The FRIS model-derived polytropic index ($\Gamma$) and temperature (T) for the selected CMEs are shown in Figures \ref{fig:gamma}a and \ref{fig:gamma}b, respectively. It describes the thermal state of the plasma without solving complex energy equations. A $\Gamma$ value less than (greater than) the adiabatic index suggests a heating state (heat-release state) for ICME plasma.  The $\Gamma$ for the selected CMEs (except CME2 and CME4) starts with a value greater than the adiabatic index ($\gamma$=1.67), suggesting a heat-releasing state. However, CME2 and CME4, even at lower heights, show a $\Gamma$ value less than the $\gamma$, indicating a heating state during their initial evolution. At greater heights, all CMEs (except CME4) remain near a $\Gamma$ value of 1, indicating an isothermal state. In contrast, CME4 approaches and maintains a value close to the adiabatic state at higher heights. This behavior may be due to the decreasing rate of the CME's propagation (or expansion) speed (Fig. \ref{fig:kine}b). With heating already present in the system, the reduction in propagation (or expansion) acceleration leads to a heat-release state for the system.

As we discussed before, the FRIS model-derived temperature values are scaled such that each CME has a temperature of $10^7 K$ at the starting point during our observation (Fig. \ref{fig:gamma}b). This will enable us to analyze the relative temperature evolution for all the selected CMEs during their evolution. The temperature for all the CMEs (except CME2 and CME6) drops rapidly at lower heights and thereafter maintains a constant value. CME2 shows a heating state ($\Gamma < 1$) throughout our model results. Thus, its temperature is not decreasing as rapidly as others. In contrast, CME4 attends a near adiabatic state at greater heights, which is reflected in its temperature profile as well. The temperature of CME4 is decreasing continuously over the observed duration. 

The analysis suggests that while successive CMEs could influence the conditions in the surrounding solar wind, the thermodynamic properties, such as the evolution of polytropic index and temperature, remain notably consistent among CMEs in this study. This result implies that CMEs may inherently possess distinct thermal characteristics, potentially set at the time of their launch, regardless of minor interaction effects with nearby solar wind alterations from preceding CMEs. Interactions between multiple CMEs within a short interval are complex and depend on various factors, including the local magnetic field vector, orientation, and relative direction of each CME. For instance, CME3, CME4, and CME5 erupted within approximately four hours. CME3 and CME4, given their longitudinal separation and faster speed of CME4, could experience a side-on interaction. Our extrapolated kinematics of CME3 and CME4 show a LE interaction height of around 50 $R_\odot$. However, taking the aspect ratio ($k$) value of 0.24 (Table \ref{tab:GCS}) and LE height ($h$) of $\approx$30 $R_\odot$ (Fig. \ref{fig:kine}c), the trailing edge height [$ h - 2(\frac{k}{1 + k} \times h)$] for CME3 is found to be $\approx$18.4 $R_\odot$. As a result, CME4 begins interacting with the trailing part of CME3 at around 20 $R_\odot$, leading to a noticeable decrease in LE acceleration beyond this height (Fig. \ref{fig:kine}b). Although CME3 and CME4 are launched in different directions, their calculated total angular width ($2\alpha + 2 \sin^{-1}k$) \citep{Thernisien2011} are 58° and 66°, respectively. Combined with the higher speed of the trailing CME4, this suggests that an interaction between CME3 and CME4 is indeed likely. A clear change in polytropic index value (Fig. \ref{fig:gamma}a) is seen for CME4 beyond $\approx$22 $R_\odot$, suggesting a thermal adjustment possibly due to the deceleration influence of interaction with CME3's trailing part. In contrast, CME4 and CME5 show minimal interaction probability, given the slower initial propagation speed of CME5. The derived thermal properties of CME5 show no significant changes within our observed remote sensing heights. The likelihood of their interaction could increase if CME4 slows down after interaction with CME3 or CME5 gains acceleration after interaction with the faster following CME6. Thus, in-situ measurements at 1 AU can give more insights into the thermal properties of CME5 and all other selected CMEs at later heights. 

In the forthcoming sections, we will analyze and disentangle the complex ICME structures using in-situ measurements. Further, we will derive the thermal state of the interacting ICME structures from in-situ measurements to get some insights into the interaction.

\begin{figure*}
   \centering
   \includegraphics[scale=0.43]{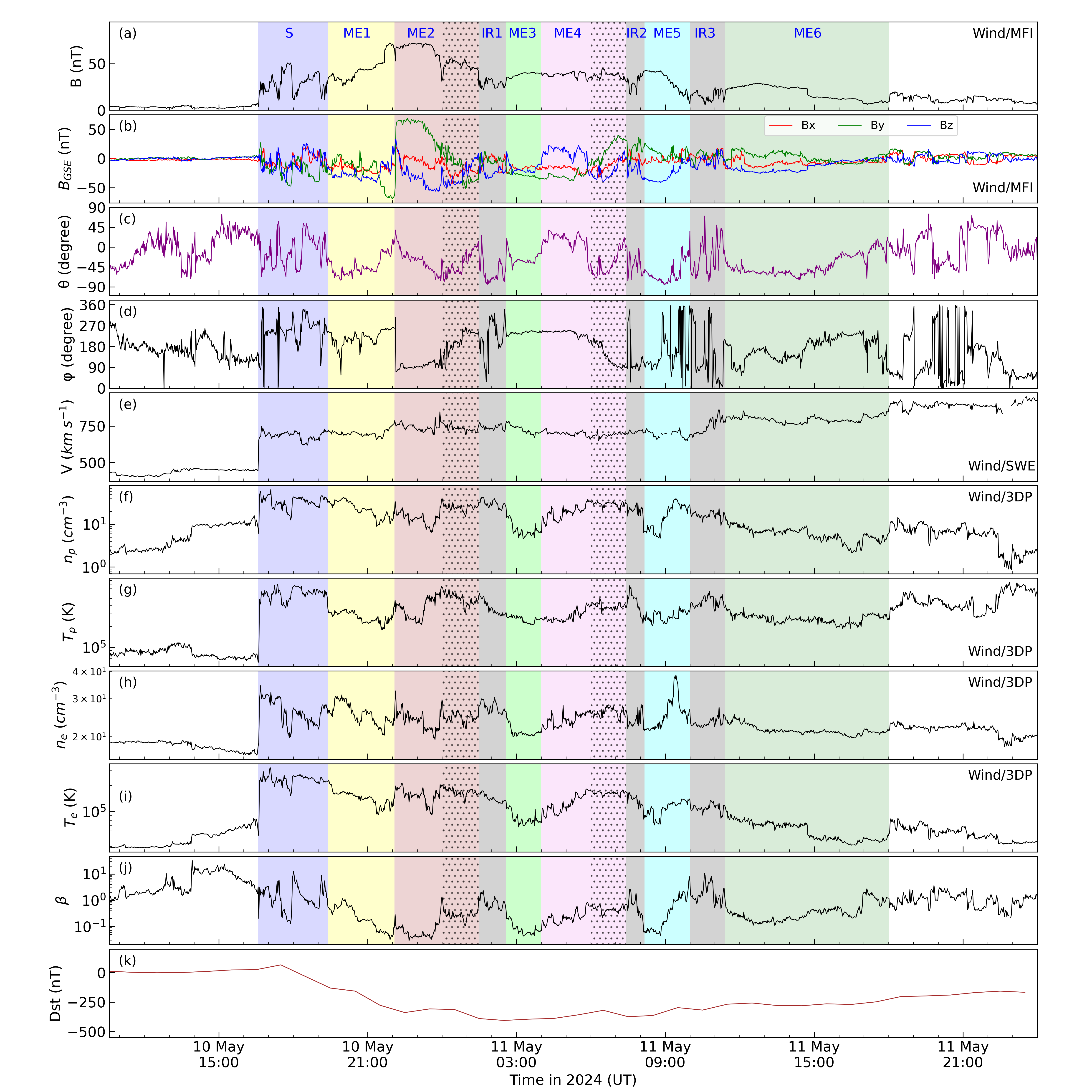}
      \caption{In-situ measurements taken by Wind spacecraft of CMEs structures driving the great geomagnetic storm on 10-11 May 2024. Panels (a) and (b) show the average magnetic field and its components in GSE coordinates, respectively. Panels (c) and (d) show the calculated inclination angle $\theta$ (with respect to the ecliptic plane) and azimuthal angle $\phi$ (0 deg pointing to the Sun) of the magnetic field. Panels (e), (f), (g), (h), (i), and (j) show the bulk solar wind speed, proton number density, proton temperature, electron number density, electron temperature, and plasma beta, respectively. Panel (k) shows the Dst index value for the selected duration of in-situ measurements. The vertical color bars in each panel show the corresponding structures, such as sheath (S), magnetic ejecta (ME), and interaction regions (IR). The dotted regions following the leading portion within ME2 and ME4 display signatures similar to double flux rope structures, which may have been inherently present during the eruption or developed later due to CME-CME interaction.}
         \label{fig:insitu}
\end{figure*}

\section{Near-Earth in-situ observations}{\label{sec:near-earth}}

We identify the large-scale solar wind structures in the in-situ observations at 1 AU associated with the candidate CMEs of the great geomagnetic storm. Figure \ref{fig:insitu} shows the solar wind properties near Earth during the geomagnetic storm observed by the Wind \citep{Ogilvie1997} spacecraft at the first Lagrange (L1) point, as well as the Dst index during 10–11 May 2024. The magnetic field data (1-minute resolution), solar wind bulk speed (92-second resolution), and plasma data (92-second resolution) are obtained from the MFI \citep{Lepping1995}, SWE \citep{Ogilvie1995}, and 3DP \citep{Lin1995} instruments onboard Wind spacecraft, respectively. The ground-based Dst index (1-hour resolution) is obtained from the World Data Center for Geomagnetism, Kyoto \citep{Nose2015}.

To aid in identifying the various substructures within the complex ejecta, we estimate the inclination angle \( \theta \) (with respect to the ecliptic plane) using the magnitude of the total magnetic field (\( B \)) and the normal component of the magnetic field (\( B_z \)) as  $\theta = \sin^{-1} \left(\frac{B_z}{B}\right)$. Since the azimuthal angle \( \phi \) rotates in the ecliptic plane (from \( 0^\circ \) to \( 360^\circ \)), it is estimated using the magnetic field components \( B_x \) and \( B_y \) as follows:

- For \( B_x > 0 \) and \( B_y > 0 \), $\phi = \tan^{-1} \left(\frac{B_y}{B_x}\right).$

- For \( B_x < 0 \) and \( B_y > 0 \), $ \phi = \tan^{-1} \left(\frac{B_y}{B_x}\right) + 180^\circ.$

- For \( B_x < 0 \) and \( B_y < 0 \), $\phi = \tan^{-1} \left(\frac{B_y}{B_x}\right) + 180^\circ.$

- For \( B_x > 0 \) and \( B_y < 0 \), $\phi = \tan^{-1} \left(\frac{B_y}{B_x}\right) + 360^\circ.$

\noindent
Such an approach to estimate the inclination and azimuthal angle of the magnetic field to identify the substructures in a single CME/MC near 1 AU is followed earlier \citep{Agarwal2024,Agarwal2025}. The values of \( \theta \) and \( \phi \) are shown in Figure \ref{fig:insitu}b and \ref{fig:insitu}c. The differently colored shaded regions correspond to various structures, including a sheath (S), interaction region (IR), and magnetic ejecta (ME), as indicated at the top of Figure \ref{fig:insitu}. The shaded regions have not been overlaid on the bottom panel of Figure \ref{fig:insitu} displaying the Dst index, as this index represents the geomagnetic response and therefore does not align with the measurements from the Wind spacecraft taken at L1.

A sudden enhancement of magnetic field (B) and speed (V) of solar wind plasma can be observed around 16:35 UT on 10 May 2024, indicating the arrival of the shock. This shock could correspond to the CME1. The arrival of shock at the bow of the magnetosphere leads to a compression of the magnetopause, which results in a rapid increase in the Earth's magnetic field on the day-side and is called a sudden storm commencement (SSC). The SSC can be seen in the Dst index profile (Fig. \ref{fig:insitu}i) of this great geomagnetic event, where the Dst index rises to a value of around 61 nT. The SSC lasted about 2 hours before we saw a negative Dst index value.

The shock was followed by a turbulent sheath region (region S in purple shade in Figure \ref{fig:insitu}) characterized by a high-value and fluctuation in magnetic field  ($B$), rapid fluctuation in magnetic field vectors ($B_{GSE}$), a rise in proton density ($n_p$), and a rise in proton temperature ($T_p$). The magnetic ejecta (ME) arrived at 19:25 UT on 10 May 2024, which we associate with CME1. This region is marked as ME1 (yellow shade) in Figure \ref{fig:insitu}. This shows a higher value of the magnetic field ($B$), rotation in magnetic field vectors ($B_{GSE}$), decrease in proton density ($n_p$), temperature ($n_p$), and plasma beta less than unity. Following ME1, we observe a sudden change in the magnetic field vectors, $\theta$, and $\phi$. Noting the smooth rotation in the magnetic field vectors and a lower plasma beta, we attribute this to ME2. Interestingly, the leading-edge (LE) speeds and accelerations of both CME1 and CME2 are estimated to be similar at 20 $R_\odot$ (Fig. \ref{fig:kine}). However, the preceding CME1 may clear the ambient medium, resulting in a lower-density region ahead of CME2, potentially reducing the drag it experiences. This enhances the likelihood of interaction between CME1 and CME2. We identified interaction regions (IRs) between MEs by analyzing in-situ magnetic and plasma parameters (gray shaded regions in Fig. \ref{fig:insitu}) as discussed in \citet{Mishra2015}. The IR1 between ME2 and ME3 is characterized by an interval of decrease in magnetic field, sudden rapid rotation in the magnetic field vector, enhancement in proton density, and plasma beta. As discussed in Sec. \ref{sec:nearsun}, CME3 has a half angle of 15$^\circ$ and source regions of longitude of 27$^\circ$ east of the Sun-Earth line, we expect a definite CME3 flank encounter at Earth ahead of CME4 and the identified region marked as ME3. Furthermore, we differentiate ME4 from ME3 based on magnetic and plasma properties, including lower proton and electron number densities, differences in magnetic field vector directions, and the $\theta$ value for ME3. We identified an IR, IR2 in between ME4 and ME5, exhibiting signatures of a decrease in the magnetic field, sudden rapid rotation in the magnetic field vector, increased proton temperature, and elevated plasma beta. We observe that IR1 exhibits a noticeable increase in proton density but does not clearly show a higher proton temperature, whereas IR2 shows an increase in proton temperature without a distinct rise in proton density. The GCS model estimated a longitude of 38$^\circ$ west of the Sun-Earth line for CME5, suggesting a possible flank encounter at Earth. However, in-situ observations of the marked region ME5 showed a smooth rotation of magnetic field vectors, lower plasma beta, and reduced proton temperature. This suggests that CME5 may have undergone a deflection toward the Sun-Earth line due to its interaction with CME6, as high-speed CMEs can realign slower CMEs toward their trajectory following an interaction. We identified an IR, IR3 in between ME5 and ME6, marked by rapid variations in magnetic field vectors, along with enhancements in density, temperature, and plasma beta.

The gray dotted regions in the trailing part of ME2 and ME4 (Fig. \ref{fig:insitu}) can be attributed to interaction-driven changes. \citet{Lugaz2013} demonstrated that the relative inclination of 90$^{\circ}$ between two ejecta increases the chances of their merging to become a single structure. The GCS model estimated the tilt angle for CME2 and CME3 to be 27$^\circ$ and 79$^\circ$, respectively. Although the relative difference in tilt is not large, the interaction with CME4 may have altered the tilt of CME3, thus enhancing the interaction between CME2 and CME3. Similarly, the tilt angles for CME4 and CME5 were estimated to be 15$^\circ$ and -83$^\circ$, respectively, which also favors their interaction. Prior to each interaction region (IR1 and IR2), the preceding ME shows noticeable deformation in magnetic field orientation and an increase in plasma parameters, including $n_p$, $T_p$, $n_e$, $T_e$, and $\beta$ (gray dotted regions in Fig. \ref{fig:insitu}).  Because of the early interaction of CME3 and CME4, the shock associated with CME4 may travel through CME3, and during the interaction of CME3 and CME2, that shock further propagates through CME2. \citet{Nlugaz2005, Lugaz2009} show that when the trailing shock impacts the leading ME, the dense sheath behind the trailing shock must remain between the two ME, even as the shock continues to propagate through the first ejecta. This phenomenon explains the presence of interaction regions between ME2 and ME3, and ME4 and ME5, as we observe in this study.  Studies utilizing numerical models \citep{Nlugaz2005, Xiong2007}, can provide a more detailed understanding of the interaction effects, which are beyond the scope of this study.

Given that CMEs are magnetized plasma structures moving through the ambient solar wind, analyzing the interaction or collision of CMEs is a complex task. Several studies have examined the nature of these collisions and the subsequent changes in CME properties, such as propagation speed, expansion, and size \citep{Temmer2014, Mishra2014, Mishra2016, FShen2016}. In this study, all CMEs except CME3 exhibit visible shock signatures in the coronagraphic images, leading us to expect that the faster CME with a shock will accelerate the preceding CME \citep{Schmidt2004, Nlugaz2005}. Thus, we noticed a gradual increase in speed for the whole combined ejecta in the in-situ measurements. We found no shock signatures associated with each faster CME, which were identified as distinct structures within the combined ejecta at 1 AU. The CME-CME interaction well before arriving at 1 AU and shock propagation through the leading magnetic ejecta might be a reason for weakening the shock properties \citep{Nlugaz2005, Lugaz2009}. In an overall view of the complex interacting structures, we observed that the speed of the combined structure continues to increase while overall, it gradually decreases in density and temperature. This indicates that the trailing magnetic ejecta continues to compress and accelerate the preceding combined structure.

\subsection{Thermodynamics using in-situ measurements}{\label{subsec:thermal_insitu}}

Some studies have been conducted to understand the thermodynamic behavior of CMEs by analyzing their in-situ observations. However, such studies are lacking for a complex ejecta formed from several interacting CMEs. By considering the CME plasma goes through a polytropic process during its evolution, the value of the polytropic index ($\Gamma$) can describe the thermal state of plasma. The polytropic equation ($T  n^{1-\Gamma}$=constant) quantifies the relationship between plasma density (n) and temperature (T). Thus, by performing a linear fit to the logarithmic values of density and temperature, the $\Gamma$ value can be determined. Previous studies have estimated the $\Gamma$ value for ICMEs using in-situ observations at distances ranging from 0.3 to 20 AU from the Sun, finding values between 1.15 and 1.33 \citep{Liu2005, Liu2006}. Model-derived results suggest a near-isothermal state in the region closer to the Sun, spanning 3–25 $R_\odot$ \citep{Khuntia2023}, implying a considerable local plasma heating inside ICME. It could be possible that because of local small-scale processes for heating, such as magnetic reconnection, turbulence, and interaction with the solar wind, the different parts of a single magnetic ejecta or complex ejecta may not be in the same thermal state at a time. In this case, a linear fit to the whole duration for a single ejecta or a complex ejecta may not give results with good correlation coefficients. Therefore, we expect that performing the linear fit to various small intervals within a complex ejecta can give a better picture of the thermal state for individual intervals. Using a similar approach, recent studies determined the radial variation of the thermal state of solar wind \citep{Nicolaou2014, Nicolaou2020}. \cite{Dayeh2022} applied this approach to analyze the statistical thermal state of various structures associated with ICMEs, revealing an adiabatic state for the pre- and post-ICME regions, while the ME exhibits a near-adiabatic heating state.

\begin{figure*}[ht]
   \centering
   \includegraphics[scale=0.43]{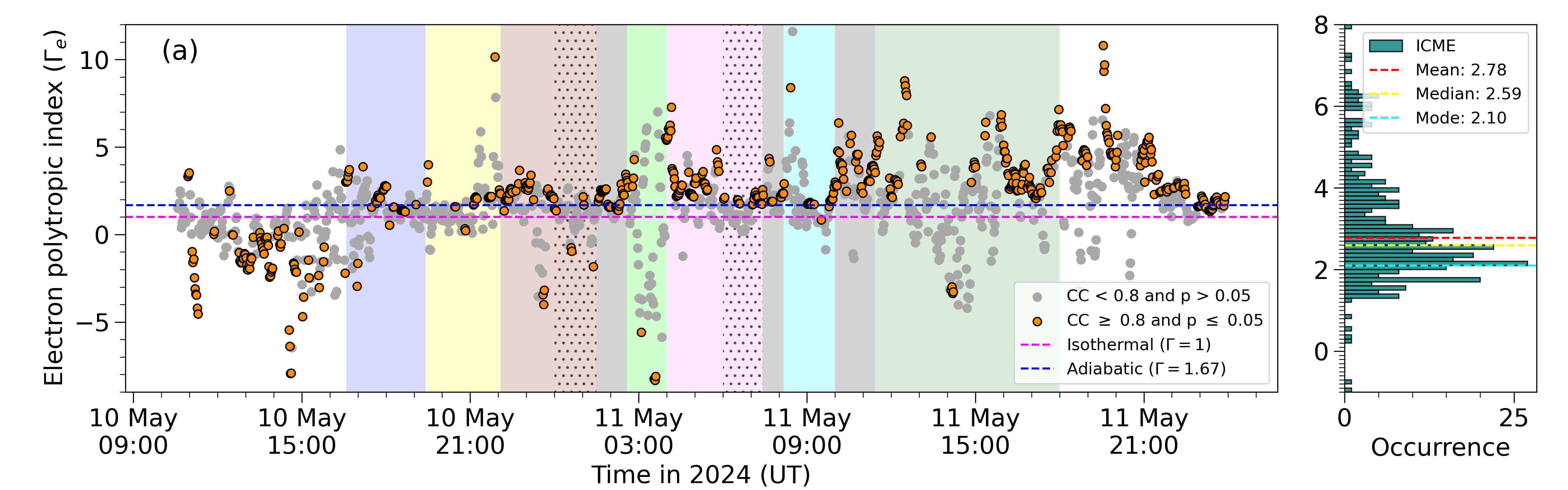}\\
   \includegraphics[scale=0.43]{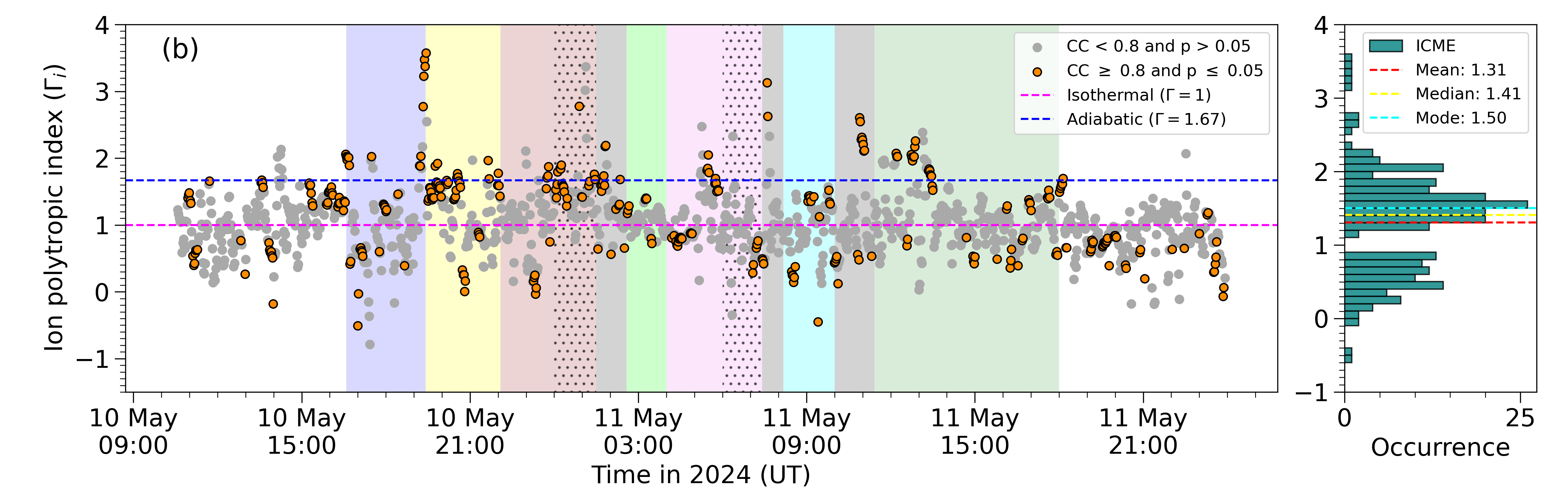}
      \caption{(a) Electron and (b) proton polytropic index for the ICME structures measured using Wind data.}
         \label{fig:insitu_thermo}
\end{figure*}

We used 92-second resolution Wind measurements for the identified complex ejecta in addition to the ambient solar wind plasma 6 hours before and after the ejecta. The pre-and post-intervals of complex ejecta are taken to compare the thermodynamics of complex ejecta with the ambient medium at that time. We used the Wind/3DP instrument observations for the plasma (both ion and electron) density and temperature, and the data is obtained from the Coordinated Data Analysis Web (CDAWeb). We selected an optimal subinterval duration of 6 data points (9.2 minutes) to fit the density and temperature variations and estimate the value of the polytropic index. We applied a linear fit between the value of log $T$ and log $n$ to calculate the value of the polytropic index for the subinterval, along with the corresponding Pearson correlation coefficient (CC) and p-value (p). The analysis was also performed using subinterval durations ranging from 4 data points (6.13 minutes) to 10 data points (15.33 minutes), with no statistically significant variation in the results. Longer subintervals may encompass a mix of different plasma streamlines, potentially leading to lower correlation coefficient values while fitting the density-temperature variations.  We repeated this fitting procedure for moving subintervals with a step size of 92 seconds. By doing so, we increased the number of data points available for $\Gamma$ measurements, improving the likelihood that the plasma within each subinterval corresponds to a single thermal state. Previous studies utilized different filtering techniques on solar wind plasma data to select the optimal subinterval for calculating $\Gamma$ \citep{Nicolaou2014, Nicolaou2020, Dayeh2022}. Our study filtered subintervals based on correlation conditions, including only those with a CC$\geq$0.8 and p$\leq$0.05. This approach ensures that the obtained gamma value accurately represents the thermal state of the entire subinterval. We did not apply the Bernoulli integral filter, which is typically used to ensure that the solar wind plasma parcel follows a single streamline. However, we consider that applying this filter is not necessary, as the correlation coefficient conditions ensure that the $\Gamma$ value reliably captures the thermal properties of the entire subinterval, even if the Bernoulli integral varies within the entire subinterval.

Figures \ref{fig:insitu_thermo}a and \ref{fig:insitu_thermo}b show the variations of the polytropic index ($\Gamma$) within the complex ejecta and within pre-and post-intervals of the complex ejecta in the solar wind medium for both electrons and ions, respectively. The background shaded regions correspond to various structures associated with the combined complex ejecta in-situ measurements, as discussed in the previous section. In each panel, we overlaid the reference lines for the adiabatic index ($\Gamma =1.67$) and isothermal index ($\Gamma =1$) values. In each panel, we plotted both the reliable (orange color) and unreliable (gray color) $\Gamma$ values obtained from fitting density and temperature. The reliable values correspond to a good fit of density and temperature to be accepted for further interpretation of the thermal state. Additionally, the right side of each panel shows the histogram for reliable $\Gamma$ values within the complete duration of complex ejecta only. We note that the electron $\Gamma$ ($\Gamma_e$) values show a clear distinction between the complex ejecta and ambient solar wind (pre and post-ejecta). The pre-interval of complex ejecta displays dominant $\Gamma_e$ values of less than 1.67, suggesting heating in these regions. In contrast, most of the measured $\Gamma_e$ values for the complex ejecta structures and post-interval are above 1.67, indicating a heat-release state. The mean and median of the occurrences of $\Gamma_e$ values inside the complex ejecta were 2.78 and 2.59, respectively. The bottom panel shows the ion $\Gamma$ ($\Gamma_i$) values for the complex ejecta and ambient solar wind (pre and post-intervals). There is a clear distinction for $\Gamma_i$ values for the complex ejecta and ambient solar wind before and after. The $\Gamma_i$ shows mostly heating signatures for pre and post-intervals of the complex ejecta. The $\Gamma_i$ values inside the complex ejecta have a two-peak distribution, showing mixtures of thermal states. Moreover, the mean and median of $\Gamma_i$ inside the complex ejecta were found to be 1.31 and 1.41, respectively. This suggests a predominant heating state within the ejecta and also significant occurrences of heat release.

Examining the thermal states of each structure within the complex ejecta reveals a trend: each subsequent ME exhibits, on average, higher $\Gamma_e$ values than its preceding ones (Fig. \ref{fig:insitu_thermo}a). This indicates that each subsequent CME in the complex ejecta, i.e., the interactions that happened later in time, that have spent a shorter time before being observed at in-situ spacecraft at 1 AU, are more intensely into heat-release states. This trend is also noticed for interacting regions IR1, IR2 and IR3. The second interaction region, IR2, having formed later than IR3, displays more pronounced heat-releasing states. We could not find a specific trend in $\Gamma_i$  values for each following interacting CME, but overall, the complex ejecta shows predominantly heating signatures, with some short intervals of heat-release states.

The $\Gamma$ value of a single CME (protons) has been statistically estimated to range between 1.1 and 1.3 from 0.3 to 20 AU \citep{Liu2006}, indicating significant heating during its evolution. On a larger scale, the $\Gamma$ value for solar wind protons typically lies between 1.5 and 1.67, while it tends to reach higher values, around 2.7 on smaller scales \citep{Nicolaou2020}. As previously discussed, the large-scale solar wind structure we're analyzing is identified as a complex ejecta and displays signatures of past interactions between potential CMEs. The difference in the electron and proton polytropic index is unsurprising as the electrons, due to their significantly smaller mass than protons, may respond more quickly and intensely to any physical processes causing thermal perturbations. 

As discussed in Section \ref{subsec:kinematics}, the CME-CME interaction heights are estimated to be well before 1 AU. These interaction heights are approximately 144 $R_\odot$ for CME1 and CME2, 54 $R_\odot$ for CME3 and CME4, and 110 $R_\odot$ for CME5 and CME6. However, it is important to note that this finding is from our simple approach, given the lack of measured kinematic information on CMEs after the coronagraphic FOV and in their post-interaction phases. This implies that establishing a one-to-one connection between CME sequences and speeds derived from remote observations with those from in-situ measurements is challenging. In in-situ observations, we could not find IRs sandwiched between ME1 and ME2 and also between ME3 and ME4 (Fig. \ref{fig:insitu}). Earlier studies have also noted the occasional absence/presence of such IRs between interacting CMEs in in-situ observations \citep{Mishra2015,Mishra2015a}. The absence of IRs between a preceding-following CME pair could be due to some characteristics of the following CME, making it less efficient in piling up ambient material and causing a smaller sheath thickness. We think that the presence/absence of IRs could also be an effect of the nature of collision between interacting CMEs; however, it is difficult to establish this because of the accurate measurements of 3D kinematics in the pre-and post-interaction phases of the CMEs.

Our analysis from remote observations shows a strong possibility of CME-CME interaction, and this is supported by the in-situ measured enhanced temperature and the reduced size of MEs identified within the complex ejecta. This shows that heating and compression happen due to CME-CME interaction, as also shown in several earlier studies \citep{Liu2012,Temmer2014,Mishra2014}. The derived polytropic index, particularly for electrons, shows a dominant heat-release state, which is possible if they are heated during these CME-CME interactions and later exhibit a heat-release state. In contrast, the heavier ions, being unable to find an equilibrium state post-interaction, still show a combination of ongoing heating and heat-release states at 1 AU. Therefore, the electron polytropic index further supports our inferences of CME-CME interaction from our extrapolated kinematics, and it could be a better proxy for CME-CME interaction than the proton polytropic index.

We also note that observations during the gray-dotted regions (discussed in Sec. \ref{sec:near-earth}) bear strong resemblances to double flux ropes (FRs) structures in ME2 and ME4. The $\phi$ values indicate that among the two flux ropes in ME2, the second flux rope shown with a dotted region is westward-directed, while the first flux rope is oriented eastward. A similar pattern is observed in ME4, where the second flux rope shown with a dotted region features an eastward-directed flux rope while the first flux rope is westward-oriented. Carefully inspecting $\theta$ values, we note that the first flux rope shows rotation from north to south, while the second tends to rotate from south to north in both ME2 and ME4. Thus, in ME2, the first and second flux ropes exhibit the orientation of NES and SWN, respectively, while in ME4, the first and second flux ropes display the orientation of NWS and SEN, respectively \citep{Bothmer1998}. We note that both flux ropes in ME2 have right-handedness, while both flux ropes in ME4 have left-handedness. This indicates that the handedness of double flux ropes within a single ME is the same. It is noted that the first flux rope in ME4 does not exhibit significant rotation in the magnetic field components, but there is an appreciable rotation in its leading portion. Therefore, it is clear that in-situ observations of magnetic fields suggest handedness similar to double flux ropes. Also, the plasma parameters, including $n_p$, $T_p$, $n_e$, $T_e$, and $\beta$, show different characteristics between two flux ropes in both ME2 and ME4. Importantly, the density in the second flux rope (shown with a dotted region) is higher than the first flux rope in both ME2 and ME4. We believe these signatures may come from CME-CME interaction, interaction between the trailing part of the ME and a following ME, resembling certain characteristics of double flux ropes. There are numerous in-situ and remote observations of CMEs suggesting the existence of multiple FRs within the same CME \citep{Ogilvie1997, Hu2003, Marubashi2007, Farrugia2011, Nieves2020, Hu2021, Wood2021}. \citet{Osherovich2013} have provided the first observational example of the presence of a double helix/double flux rope in an erupting prominence and in-situ measurements of an ICME. Similarly, \cite{Hu2003, Hu2021} also identified the presence of a double flux rope structure with opposite field polarities using the Grad-Shafranov (GS) reconstruction technique on observed ICMEs.

We focus on the thermal state of MEs showing distinct characteristics akin to double flux rope structures. We note different mean values of $\Gamma_e$ within the two flux ropes of ME2 ($\Gamma_e$ = 1.9 for the first flux rope and 0.4 for the second) and ME4 ($\Gamma_e$ = 3.0 for the first flux rope and 2.0 for the second), as shown in Figure \ref{fig:insitu_thermo}a. The lower value of the polytropic index during the second flux rope, with higher density, in both ME2 and ME4, indicates a weak correlation between temperature and density. In fact, the temperature and density between the first and second flux ropes of ME2 are oppositely correlated. In addition to plasma and magnetic field observations during ME2 and ME4, the thermal states also exhibit a resemblance with double flux rope structures. Our finding is consistent with the study of \citet{Osherovich2013}, which also suggests that the two flux ropes can exhibit distinct electron polytropic indices. We cannot rule out the possibility that these structures merely resemble double flux ropes but actually result solely from CME-CME interactions rather than originating as a double flux rope system near the Sun. Identifying the physical processes responsible for the formation of these double flux ropes, particularly within this complex ejecta, is beyond the scope of this study. A detailed investigation combining remote sensing, in-situ observations, and modeling of an isolated event could offer deeper insights into such structures.

\subsection{Comparison of FRIS-model and in-situ derived thermal states}{\label{subsec:thermal_fris_insitu}}

The FRIS model-derived polytropic index ($\Gamma$) was calculated (Sec. \ref{subsec:thermal}) under a polytropic approximation for the entire CME with an average temperature, $T=(T_e + T_p )/2$ and a number density, $n=n_e=n_p $. Therefore, an effective polytropic index, combining both electrons and proton polytropic index of the entire ME from in-situ at 1 AU, needs to be compared with the model-derived polytropic index of the CME near the Sun. Since \(T_e\) is associated with \(\Gamma_e\) and \(T_p\) with \(\Gamma_p\), we can calculate the effective polytropic index \(\Gamma_{eff}\) as the weighted average of \(\Gamma_e\) and \(\Gamma_p\) and can be expressed as
\[\Gamma_{eff} \approx \frac{\Gamma_e T_e + \Gamma_p T_p}{T_e + T_p}\] where the weights are proportional to the temperatures of the electron and proton populations. In our study, we assume that the mean value of estimated \(\Gamma_e\), \(\Gamma_p\), $T_e$, and $T_p$ corresponding to different chosen intervals within a ME represent the thermal state of that ME. The mean values representing the thermal state at 1 AU are listed in Appendix Table \ref{tab:eff_gamma}. Using these mean values, we further calculated $\Gamma_{eff}$ for ME1, ME2, ME3, ME4, ME5, and ME6, which are 1.77, 1.47, 0.8, 1.63, 1.26, and 1.75, respectively. ME1 and ME6 exhibit a heat-release state at 1 AU, ME2 and ME4 indicate a near-adiabatic heating state, while ME3 and ME5 display a near-isothermal heating state.

On comparing the estimates of $\Gamma_{eff}$ near 1 AU with those $\Gamma$ derived from the model, we find a large difference for ME1 and ME6, almost no change in ME3, ME4, and ME5, and a moderate change for ME2. Such a direct comparison based on only two-point measurements (one close to the Sun and the other at 1 AU) to understand any change in the thermal states of CMEs due to interaction could have been meaningful for only one interacting CME pair. However, in our case, there are multiple interacting CME pairs, and they are expected to undergo multiple heating and heat-release states. Therefore, a one-to-one comparison between the in-situ and model-derived $\Gamma$ would be extremely difficult to achieve a meaningful thermal history. Our study emphasizes that the thermodynamic evolution of interacting CMEs is complex; therefore, one needs information on the continuous thermal history of CMEs during their pre-, ongoing, and post-interaction phases for better understanding. Future studies utilizing HIs observations combined with in-situ observations of CMEs at distances covering their pre- and post-interaction phases would provide more insights into their ongoing thermal states.

%%%-----------------------------------------------------------------

\section{Conclusions}{\label{sec:conclusions}}
In this study, we identify a series of six CMEs ejected in succession from the Sun on 8-9 May 2024 that led to the great geomagnetic storm with its onset on 10 May 2024. Using data from SOHO/LASCO-C2, STEREO-A/COR2, and SDO/AIA, we analyzed the identified CMEs and applied the GCS model to derive their 3D kinematics and understand their interactions before arriving on Earth as a complex ejecta. Our study also focused on the thermodynamics of the selected CMEs using both remote and in-situ observations. The following points summarize the findings of our study,

   \begin{enumerate}
      
      \item Using the measured 3D kinematics, we conclude the potential interactions among the selected CMEs (CME1 and CME2 at 144 R$_\odot$, CME3 and CME4 at 54 R$_\odot$, CME5 and CME6 at 110 R$_\odot$) to form a complex ejecta before reaching 1 AU.

      \item The study also identified different MEs corresponding to near-Sun identified CMEs in the in-situ observations and noted the signatures of the CME-CME interaction. The interacting CMEs show clear signatures of heating and compression. This suggests that CME interactions significantly influence their large-scale properties and impact on space weather near Earth.

      \item We identified IRs between pairs of MEs in in-situ observations, indicating that CME-CME interactions occurred as they propagated toward Earth. Additionally, we observed a single ME (ME2 and ME4) displaying magnetic field, polytropic index, and density-temperature characteristics similar to double flux rope structures. Such double flux rope-like structures may appear due to variations in ME properties resulting from CME-CME interactions.

      \item The thermal evolution of CMEs even before the interaction varies significantly, with most CMEs transitioning to an isothermal state at higher coronal heights, while exceptions like CME4 approach an adiabatic state. The FRIS model shows that heating or heat release in the plasma depends on the CME's propagation and expansion, with slower expansion leading to heat release, as observed in CME4. We did not notice a difference in CME's thermodynamics in their later phase at coronal heights despite the fact that each of them is traveling in a different pre-conditioned medium.

      \item In-situ observations at 1 AU show that electrons within the interacting CMEs forming complex ejecta exhibit a predominant heat-release state, as indicated by their polytropic index values ($\Gamma_e$ > 1.67). We also note that the post-interval of the complex ejecta shows a predominant heat release while the pre-interval of the complex ejecta shows a heating state, suggesting distinct thermal processes occurring inside and outside the interacting large-scale structures.

      \item The polytropic index for ions ($\Gamma_i$) inside the complex ejecta shows a significant bimodal distribution, with a predominant heating state with some intervals of heat-release states, which is unlike the electron polytropic index. The difference in the electron and proton polytropic index could be due to a more gradual thermal evolution of protons than electrons in interacting CMEs, and the thermal state can depend on the duration they have spent post-interaction.

   \end{enumerate}

Overall, the combination of thermal and kinematic analyses offers valuable insights into the evolution of the interacting CMEs. The pairs of interacting CMEs have caused the great geomagnetic storm, which was not the primary focus of the study; it highlights the broader implications of such interactions in influencing CME plasma properties and their potential geo-effectiveness. Although our study confirms the CME-CME interactions from both remote and in-situ observations focusing on the thermodynamic evolution of CMEs, we plan to use heliospheric imaging observations to provide further insights into their pre- and post-interaction behaviors. Noticing properties of different substructures, such as IRs and double FRs within the complex ejecta, requires further in-depth investigation of them using remote and in-situ observations to better examine their roles in causing such a rare great geomagnetic storm.

\begin{acknowledgements}
      We appreciate the anonymous referee's valuable comments, insightful questions, and constructive suggestions that greatly improved this manuscript.
\end{acknowledgements}

%%%%%--------------------------------------------------------------------

%%%%%--------------------------------------------------------------------

\begin{appendix}

\onecolumn
\section{The FRIS model-fitting errors}
  \begin{figure*}[ht]
   \centering
   \includegraphics[scale=0.32,trim={1cm 1cm 1cm 0cm}, clip]{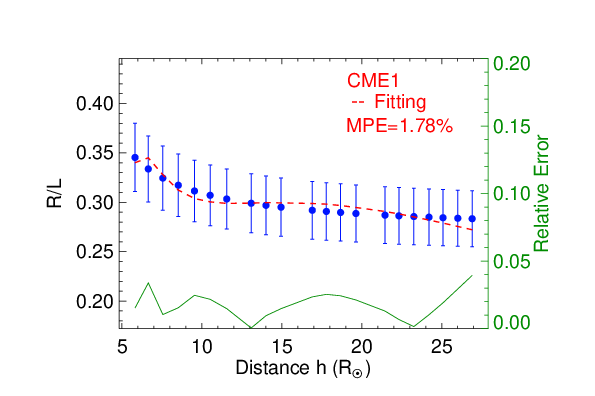}
   \includegraphics[scale=0.32,trim={1cm 1cm 1cm 0cm}, clip]{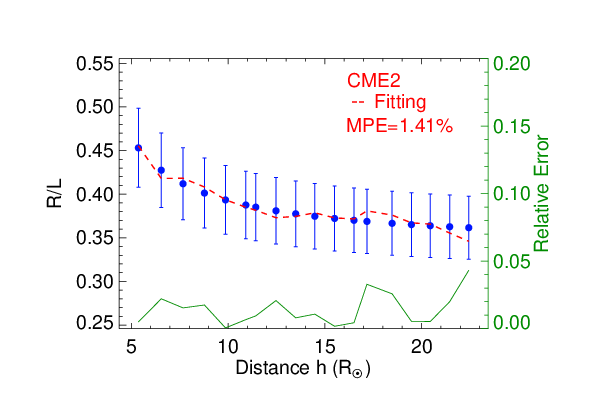}
    \includegraphics[scale=0.32,trim={1cm 1cm 1cm 0cm}, clip]{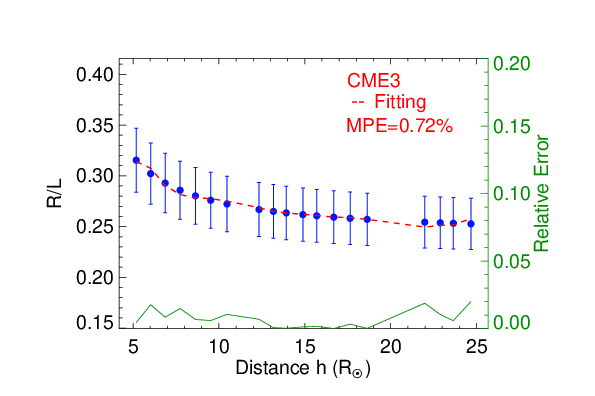}\\
     \includegraphics[scale=0.32,trim={1cm 1cm 1cm 0cm}, clip]{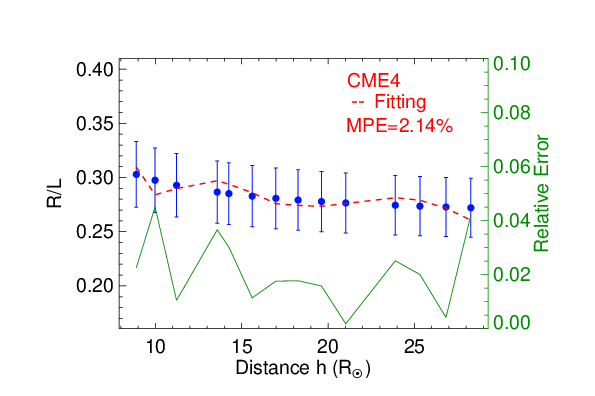} 
     \includegraphics[scale=0.32,trim={1cm 1cm 1cm 0cm}, clip]{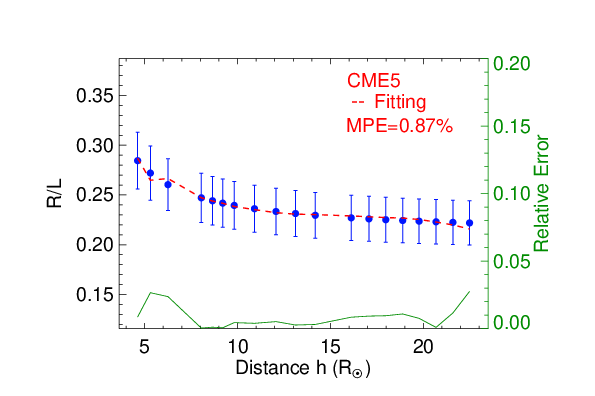}
      \includegraphics[scale=0.32,trim={1cm 1cm 1cm 0cm}, clip]{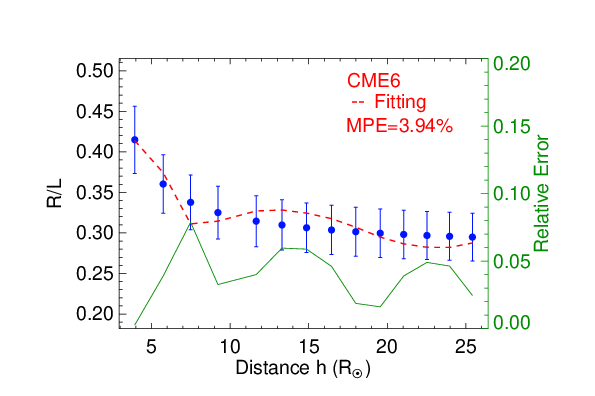}
      \caption{Model-fitting errors for the selected fast-speed CMEs. Blue dots: the left-hand side of Equation (1), Red-dash line: the right-hand side of Equation (1) in \citet{Khuntia2023}, and green line: relative fitting errors. MPE: mean percentage error.}
         \label{fig:FRIS_error}
   \end{figure*}
\twocolumn

\onecolumn
\section{Mean values of in-situ plasma parameters}
\begin{table*}[ht]
\centering
\caption{Mean values of plasma parameters for different magnetic ejecta and the calculated effective polytropic index ($\Gamma_{eff}$) from in-situ measurements at 1 AU.}
\label{tab:eff_gamma}
\begin{tabular}{|c|c|c|c|c|c|}
\hline
\multirow{2}{*}{Magnetic Ejecta No.} & \multicolumn{4}{c|}{Mean values for each magnetic ejecta} & \multirow{2}{*}{$\Gamma_{eff}$} \\  
\cline{2-5}
& $T_e$ ($10^5$ K) \rule{0pt}{2.2ex} & $T_p$ ($10^5$ K) & $\Gamma_e$ & $\Gamma_p$ & \\
\hline
ME1 & 1.46  & 2.74  & 2.42 & 1.43  & 1.77  \\
ME2 & 1.62  & 4.35  & 1.70  & 1.38  & 1.47  \\
ME3 & 0.83 & 2.5  & -0.31  & 1.17 & 0.8  \\
ME4 & 1.41 & 3.24 & 2.96  & 1.05 & 1.63  \\
ME5 & 1.1 & 1.81  & 2.33  & 0.85  & 1.26  \\
ME6 & 0.61 & 2.56 & 3.74 & 1.27 & 1.75 \\
\hline
\end{tabular}
\end{table*}
\twocolumn

 \end{appendix}

%%%%%--------------------------------------------------------------------
\end{document}